\documentclass[aps,prl,superscriptaddress,twocolumn,10pt]{revtex4-1}

\usepackage{graphicx}
\usepackage{physics}
\usepackage{algpseudocode}
\usepackage{pythonhighlight}
\usepackage{amsmath,amssymb,amsfonts}
\usepackage{hyperref}
\usepackage{xcolor}
\usepackage{tikz}
\usetikzlibrary{positioning,arrows.meta,fit,backgrounds}
\usepackage{fontawesome5}

\definecolor{good}{RGB}{30,132,73}
\definecolor{bad}{RGB}{192,57,43}
\definecolor{ink}{RGB}{43,43,43}
\definecolor{mut}{RGB}{90,90,90}
\definecolor{rowblue}{RGB}{233,242,250}
\definecolor{rowred}{RGB}{250,236,234}
\definecolor{rowgreen}{RGB}{236,247,238}
\definecolor{watercol}{RGB}{52,152,219}
\definecolor{boltcol}{RGB}{243,156,18}
\definecolor{moneycol}{RGB}{30,132,73}
\newcommand{\dataset}[4]{%
  \def\hw{0.85}\def\hh{0.8}
  \begin{scope}
    \fill[white,draw=gray!60,line width=0.8pt,rounded corners=2pt]
      (#1-\hw,#2-\hh) rectangle (#1+\hw,#2+\hh);
    \pgfmathsetmacro{\pad}{0.08}
    \pgfmathsetmacro{\cw}{(2*\hw-2*\pad)/#4}
    \pgfmathsetmacro{\ch}{(2*\hh-2*\pad)/#4}
    \foreach \i in {0,...,\numexpr#4-1\relax}{
      \foreach \j in {0,...,\numexpr#4-1\relax}{
        \fill[blue!8,draw=blue!18,line width=0.4pt]
          (#1-\hw+\pad+\i*\cw+0.02,#2-\hh+\pad+\j*\ch+0.02)
          rectangle (#1-\hw+\pad+\i*\cw+\cw-0.02,#2-\hh+\pad+\j*\ch+\ch-0.02);
      }
    }
    \node[font=\scriptsize,mut,anchor=north] at (#1,#2-\hh-0.08) {#3};
  \end{scope}
}
\newcommand{\outcome}[4]{%
  \begin{scope}
    \ifnum\pdfstrcmp{#3}{good}=0
      \colorlet{oc}{good}\colorlet{ocfill}{rowgreen}\def\ocsym{\faCheckCircle}
    \else
      \colorlet{oc}{bad}\colorlet{ocfill}{rowred}\def\ocsym{\faTimesCircle}
    \fi
    \fill[ocfill,draw=oc,line width=1.3pt,rounded corners=6pt]
      (#1-2.55,#2-0.46) rectangle (#1+2.55,#2+0.46);
    \node[oc,font=\large] at (#1-2.2,#2) {\ocsym};
    \node[oc,font=\small\bfseries,anchor=west] at (#1-1.95,#2) {#4};
  \end{scope}
}
\newcommand{\deepnet}[5]{%
  \begin{scope}
  \foreach \li [count=\lc from 0] in {5,6,6,6,2}{
    \pgfmathsetmacro{\lx}{#1 + (#2-#1)*\lc/4}
    \ifnum\lc=4
      \coordinate (D4-1) at (\lx,#3+0.42);
      \coordinate (D4-2) at (\lx,#3-0.42);
    \else
      \foreach \ni in {1,...,\li}{
        \pgfmathsetmacro{\ny}{#3 + #4 - (\ni-1)*2*#4/(\li-1)}
        \coordinate (D\lc-\ni) at (\lx,\ny);}
    \fi
  }
  \foreach \lc [evaluate=\lc as \nc using int(\lc+1)] in {0,1,2,3}{
    \pgfmathtruncatemacro{\liA}{{5,6,6,6,2}[\lc]}
    \pgfmathtruncatemacro{\liB}{{5,6,6,6,2}[\nc]}
    \foreach \a in {1,...,\liA}{\foreach \b in {1,...,\liB}{
      \draw[gray!55,line width=0.2pt] (D\lc-\a) -- (D\nc-\b);}}
  }
  \foreach \lc in {0,1,2,3,4}{
    \pgfmathtruncatemacro{\li}{{5,6,6,6,2}[\lc]}
    \foreach \ni in {1,...,\li}{
      \fill[white,draw=black,line width=0.6pt] (D\lc-\ni) circle (#5);}
  }
  \end{scope}
}
\newcommand{\uv}[1]{{\fontsize{5.2}{5.6}\selectfont\ensuremath{#1}}}
\def\DSX{1.35}\def\ALx{2.35}\def\ARRLEN{0.8}\def\ARx{8.25}\def\OUTX{11.9}
\def\Wl{-0.3}\def\Wr{15.1}
\usepackage{tikz}
\usetikzlibrary{positioning,arrows.meta}
\usepackage{quantikz}

\definecolor{convcol}{RGB}{173,216,230}
\definecolor{poolcol}{RGB}{152,223,152}
\definecolor{densecol}{RGB}{255,241,150}
\definecolor{inputcol}{RGB}{210,210,210}
\definecolor{paramcol}{RGB}{90,90,90}
\def\parbase{-2.4}
\newcommand{\fmap}[9]{%
  \pgfmathsetmacro{\dx}{0.62*#4}%
  \pgfmathsetmacro{\dy}{0.50*#4}%
  \pgfmathsetmacro{\yb}{#2-#3/2}%
  \pgfmathsetmacro{\yt}{#2+#3/2}%
  \pgfmathsetmacro{\xr}{#1+#3}%
  \fill[#5!75!black] (\xr,\yb) -- (\xr,\yt) -- (\xr+\dx,\yt+\dy) -- (\xr+\dx,\yb+\dy) -- cycle;
  \draw[line width=0.3pt] (\xr,\yb) -- (\xr,\yt) -- (\xr+\dx,\yt+\dy) -- (\xr+\dx,\yb+\dy) -- cycle;
  \fill[#5!55!white] (#1,\yt) -- (\xr,\yt) -- (\xr+\dx,\yt+\dy) -- (#1+\dx,\yt+\dy) -- cycle;
  \draw[line width=0.3pt] (#1,\yt) -- (\xr,\yt) -- (\xr+\dx,\yt+\dy) -- (#1+\dx,\yt+\dy) -- cycle;
  \fill[#5] (#1,\yb) rectangle (\xr,\yt);
  \draw[line width=0.3pt] (#1,\yb) rectangle (\xr,\yt);
  \node[anchor=west, font=\scriptsize] at (#1+0.07,#2) {#7};
  \node[anchor=north, font=\scriptsize] at (#1+#3/2,\yb-0.04) {#6};
  \node[anchor=south, font=\scriptsize] at (#1+#3/2+\dx,\yt+\dy+0.02) {#8};
  \node[anchor=north, font=\scriptsize\itshape, paramcol] at (#1+#3/2,\parbase) {#9};
}
\newcommand{\oplabel}[2]{\node[anchor=south, font=\small] at (#1,1.95) {#2};}
\newsavebox{\embedbox}
\savebox{\embedbox}{\begin{tikzpicture}[baseline=(lbl.base)]
  \fill[inputcol,draw=black,line width=0.4pt] (0,0) rectangle (1.0,1.0);
  \foreach \i in {1,...,7}{%
    \draw[gray!55,line width=0.15pt] (\i*0.125,0)--(\i*0.125,1.0);
    \draw[gray!55,line width=0.15pt] (0,\i*0.125)--(1.0,\i*0.125);
  }
  \node[font=\scriptsize] at (0.5,-0.20) {$8\times8$};
  \draw[-{Stealth},line width=0.5pt] (0.5,-0.36) -- (0.5,-0.66);
  \node[font=\footnotesize] (lbl) at (0.5,-0.88) {Amp.\ Embed.};
\end{tikzpicture}}
\definecolor{mygreen}{RGB}{0,120,0}
\usepackage{natbib}
\usepackage{changes}
\usepackage[english]{babel}
\usepackage{subfiles}
\usepackage{subcaption}
\usepackage{float}
\usepackage{siunitx}
\usepackage{mdframed}


\AtBeginDocument{\RenewCommandCopy\qty\SI}
\begin{document}

\title{Experimental evidence of generalization in quantum machine learning \\in small-data regime}
\author{Leena Anthony}
\affiliation{School of Life Sciences, University of Applied Sciences Northwestern Switzerland (FHNW), CH-4132 Muttenz, Switzerland}
\affiliation{\'{E}cole Polytechnique F\'{e}d\'{e}rale de Lausanne (EPFL), CH-1015 Lausanne, Switzerland}
\author{Artemiy Burov}
\affiliation{School of Life Sciences, University of Applied Sciences Northwestern Switzerland (FHNW), CH-4132 Muttenz, Switzerland}
\affiliation{\'{E}cole Polytechnique F\'{e}d\'{e}rale de Lausanne (EPFL), CH-1015 Lausanne, Switzerland}
\author{Nicolas Piro}
\affiliation{MatterDecoder, CH-3008 Bern, Switzerland}
\author{Matteo Dal Peraro}
\affiliation{\'{E}cole Polytechnique F\'{e}d\'{e}rale de Lausanne (EPFL), CH-1015 Lausanne, Switzerland}
\author{Cl\'{e}ment Javerzac}
\email{clement.javerzac@fhnw.ch}
\affiliation{School of Life Sciences, University of Applied Sciences Northwestern Switzerland (FHNW), CH-4132 Muttenz, Switzerland}
\affiliation{\'{E}cole Polytechnique F\'{e}d\'{e}rale de Lausanne (EPFL), CH-1015 Lausanne, Switzerland}
\affiliation{MatterDecoder, CH-3008 Bern, Switzerland}

\begin{abstract}
Quantum machine learning has emerged as a promising paradigm for learning from
limited data, a central bottleneck in domains such as medical imaging, clinical
trials, and rare diseases. Quantum convolutional neural networks (QCNNs) are
particularly attractive in this setting, combining a hierarchical architecture
with strong inductive bias and a parameter count that grows only logarithmically
with system size. Their appeal rests in part on the generalization bounds of
\citet{caro2022generalization}, which show that the generalization error of a
quantum model scales with the number of trainable parameters rather than with the
Hilbert-space dimension, placing QCNNs in a potentially sample-efficient regime.
We develop a hardware-compatible QCNN with mid-circuit measurement and classical
feed-forward, and show on a binary handwritten-digit task that strong test
performance is achievable from as few as \(10\) training samples, with the
generalization error decreasing as the training set grows. At a matched
45-parameter budget the QCNN learns where an equally small classical convolutional
network stays at chance, although an unconstrained classical baseline with roughly
\(25{,}000\) parameters remains strongest when data are plentiful. Transpiling
amplitude and angle encoded circuits across image resolutions from \(2\times2\)
to \(512\times512\) pixels then exposes the dominant scaling bottleneck: amplitude
encoding stays qubit-efficient but grows extremely deep, whereas angle encoding
stays shallow but becomes qubit-prohibitive. Finally, on the medically motivated
BreastMNIST benchmark the QCNN does not surpass the unconstrained classical
network, yet it learns consistently above chance using orders of magnitude fewer
parameters. Our results indicate that for QCNNs, learning from few samples is
attainable in practice, whereas scaling to realistic image data is constrained
less by optimization than by data encoding and hardware execution.
\end{abstract}

\maketitle

\section{Introduction\label{Introduction}}

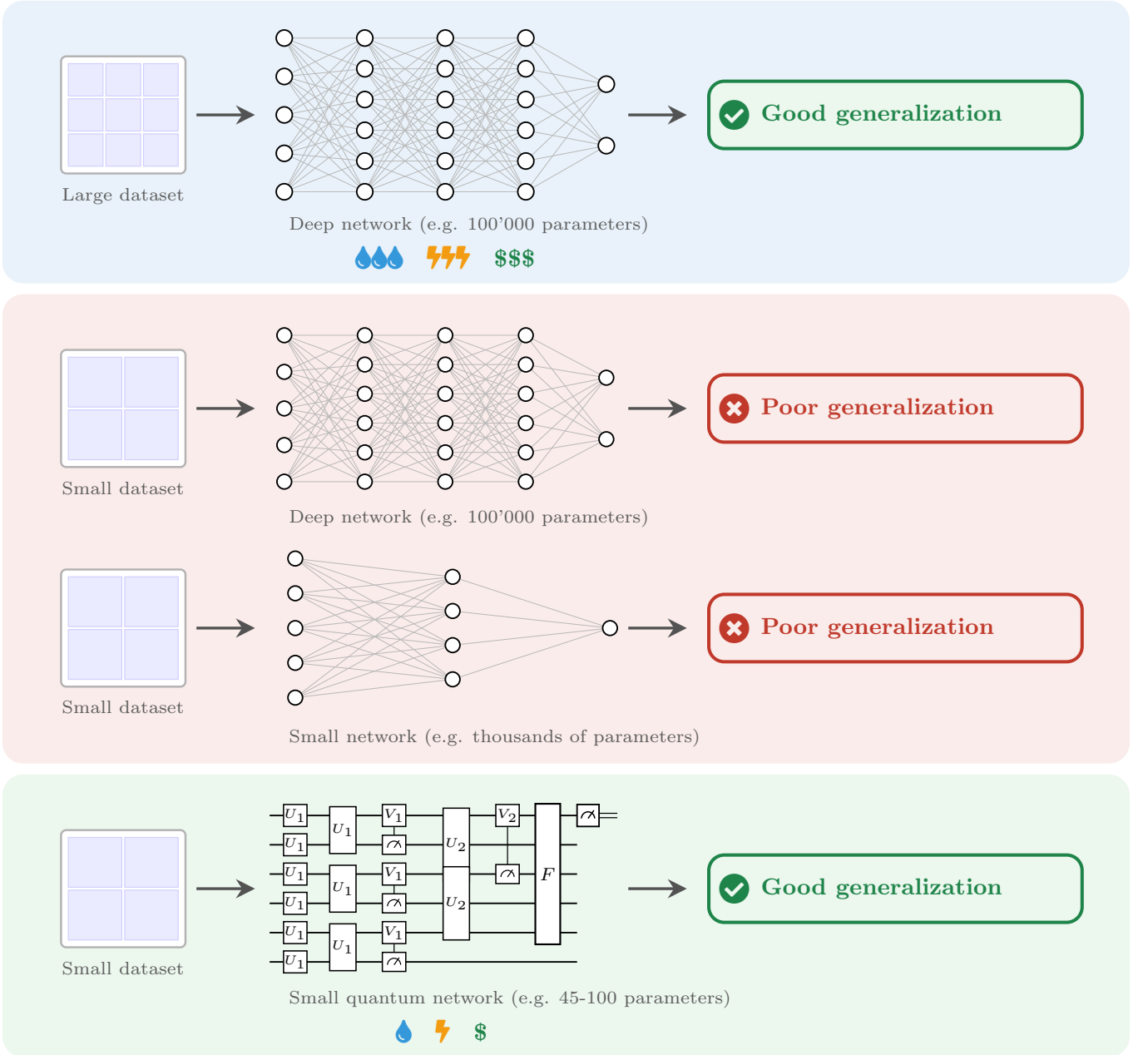
\begin{figure*}[t]
  \centering
  \resizebox{\linewidth}{!}{%
\begin{tikzpicture}[x=1cm,y=1cm,
  every node/.style={font=\sffamily},
  arr/.style={-{Stealth[length=2.6mm,width=2.6mm]}, line width=1.1pt, gray!65!black}]
 
\def\BlueT{1.56}\def\BlueB{-2.30}
\def\RedT{-2.45}\def\RedB{-8.86}
\def\GreenT{-9.01}\def\GreenB{-12.87}
 
\def\pA{0}
\begin{scope}[on background layer]
  \fill[rowblue,rounded corners=8pt] (\Wl,\BlueT) rectangle (\Wr,\BlueB);
\end{scope}
\dataset{\DSX}{\pA}{Large dataset}{3}
\draw[arr] (\ALx,\pA) -- (\ALx+\ARRLEN,\pA);
\deepnet{3.55}{7.95}{\pA}{1.05}{0.11}
\node[font=\scriptsize,mut,anchor=west] at (3.5,\pA-1.5) {Deep network (e.g. 100'000 parameters)};
\node[font=\small] at (5.75,\pA-1.95) {%
  \textcolor{watercol}{\faTint\faTint\faTint}\quad
  \textcolor{boltcol}{\faBolt\faBolt\faBolt}\quad
  \textcolor{moneycol}{\textbf{\$\$\$}}};
\draw[arr] (\ARx,\pA) -- (\ARx+\ARRLEN,\pA);
\outcome{\OUTX}{\pA}{good}{Good generalization}
 
\def\pBu{-4.01}
\def\pBl{-7.01}
\begin{scope}[on background layer]
  \fill[rowred,rounded corners=8pt] (\Wl,\RedT) rectangle (\Wr,\RedB);
\end{scope}
 
\dataset{\DSX}{\pBu}{Small dataset}{2}
\draw[arr] (\ALx,\pBu) -- (\ALx+\ARRLEN,\pBu);
\deepnet{3.55}{7.95}{\pBu}{1.0}{0.10}
\node[font=\scriptsize,mut,anchor=west] at (3.5,\pBu-1.5) {Deep network (e.g. 100'000 parameters)};
\draw[arr] (\ARx,\pBu) -- (\ARx+\ARRLEN,\pBu);
\outcome{\OUTX}{\pBu}{bad}{Poor generalization}
 
\dataset{\DSX}{\pBl}{Small dataset}{2}
\draw[arr] (\ALx,\pBl) -- (\ALx+\ARRLEN,\pBl);
\def\sxL{3.7}\def\sxM{5.85}\def\sxR{8.0}
\begin{scope}
  \foreach \ni in {1,...,5}{\pgfmathsetmacro{\ny}{\pBl+0.95-(\ni-1)*1.9/4}\coordinate (s0-\ni) at (\sxL,\ny);}
  \foreach \ni in {1,...,4}{\pgfmathsetmacro{\ny}{\pBl+0.7-(\ni-1)*1.4/3}\coordinate (s1-\ni) at (\sxM,\ny);}
  \coordinate (s2-1) at (\sxR,\pBl);
  \foreach \a in {1,...,5}{\foreach \b in {1,...,4}{\draw[gray!55,line width=0.25pt] (s0-\a)--(s1-\b);}}
  \foreach \b in {1,...,4}{\draw[gray!55,line width=0.25pt] (s1-\b)--(s2-1);}
  \foreach \ni in {1,...,5}{\fill[white,draw=black,line width=0.6pt] (s0-\ni) circle (0.10);}
  \foreach \ni in {1,...,4}{\fill[white,draw=black,line width=0.6pt] (s1-\ni) circle (0.10);}
  \fill[white,draw=black,line width=0.6pt] (s2-1) circle (0.10);
\end{scope}
\node[font=\scriptsize,mut,anchor=west] at (3.5,\pBl-1.5) {Small network (e.g. thousands of parameters)};
\draw[arr] (\ARx,\pBl) -- (\ARx+\ARRLEN,\pBl);
\outcome{\OUTX}{\pBl}{bad}{Poor generalization}
 
\def\pC{-10.57}
\begin{scope}[on background layer]
  \fill[rowgreen,rounded corners=8pt] (\Wl,\GreenT) rectangle (\Wr,\GreenB);
\end{scope}
\dataset{\DSX}{\pC}{Small dataset}{2}
\draw[arr] (\ALx,\pC) -- (\ALx+\ARRLEN,\pC);
 
\begin{scope}[shift={(0,\pC)}, every path/.style={black}]
  \def\xEnd{7.55}
  \foreach \k in {1,...,6}{\pgfmathsetmacro{\wy}{1.0-(\k-1)*2.0/5}
    \draw[line width=0.6pt] (3.35,\wy) -- (\xEnd,\wy);}
  \def\cUa{3.7}
  \foreach \k in {1,...,6}{\pgfmathsetmacro{\wy}{1.0-(\k-1)*2.0/5}
    \draw[fill=white,line width=0.5pt] (\cUa-0.16,\wy-0.15) rectangle (\cUa+0.16,\wy+0.15);
    \node at (\cUa,\wy) {\uv{U_1}};}
  \def\cUb{4.35}
  \foreach \p in {0,1,2}{
    \pgfmathsetmacro{\ya}{1.0-(2*\p)*2.0/5}\pgfmathsetmacro{\yb}{1.0-(2*\p+1)*2.0/5}
    \pgfmathsetmacro{\ym}{(\ya+\yb)/2}
    \draw[fill=white,line width=0.5pt] (\cUb-0.18,\ym-0.32) rectangle (\cUb+0.18,\ym+0.32);
    \node at (\cUb,\ym) {\uv{U_1}};}
  \def\cV{5.05}
  \foreach \p in {0,1,2}{
    \pgfmathsetmacro{\ya}{1.0-(2*\p)*2.0/5}\pgfmathsetmacro{\yb}{1.0-(2*\p+1)*2.0/5}
    \draw[fill=white,line width=0.5pt] (\cV-0.16,\ya-0.15) rectangle (\cV+0.16,\ya+0.15);
    \node at (\cV,\ya) {\uv{V_1}};
    \draw[fill=white,line width=0.5pt] (\cV-0.16,\yb-0.13) rectangle (\cV+0.16,\yb+0.13);
    \draw[line width=0.4pt] (\cV-0.09,\yb-0.04) arc (180:0:0.09);
    \draw[line width=0.4pt] (\cV,\yb-0.04) -- (\cV+0.07,\yb+0.06);
    \draw[line width=0.4pt] (\cV,\yb+0.13) -- (\cV,\ya-0.15);}
  \def\cU2{5.9}
  \foreach \ym in {0.6,-0.2}{
    \draw[fill=white,line width=0.5pt] (\cU2-0.18,\ym-0.5) rectangle (\cU2+0.18,\ym+0.5);
    \node at (\cU2,\ym) {\uv{U_2}};}
  \def\cV2{6.6}
  \pgfmathsetmacro{\yA}{1.0}\pgfmathsetmacro{\yC}{1.0-2*2.0/5}\pgfmathsetmacro{\yE}{1.0-4*2.0/5}
  \draw[fill=white,line width=0.5pt] (\cV2-0.16,\yA-0.15) rectangle (\cV2+0.16,\yA+0.15);
  \node at (\cV2,\yA) {\uv{V_2}};
  \draw[fill=white,line width=0.5pt] (\cV2-0.16,\yC-0.13) rectangle (\cV2+0.16,\yC+0.13);
  \draw[line width=0.4pt] (\cV2-0.09,\yC-0.04) arc (180:0:0.09);
  \draw[line width=0.4pt] (\cV2,\yC-0.04) -- (\cV2+0.07,\yC+0.06);
  \draw[line width=0.4pt] (\cV2,\yC+0.13) -- (\cV2,\yA-0.15);
  \def\cF{7.15}
  \draw[fill=white,line width=0.7pt] (\cF-0.17,\yE-0.16) rectangle (\cF+0.17,\yA+0.16);
  \node[font=\scriptsize] at (\cF,{(\yA+\yE)/2}) {$F$};
  \draw[fill=white,line width=0.6pt] (\xEnd,\yA-0.15) rectangle (\xEnd+0.30,\yA+0.15);
  \draw[line width=0.5pt] (\xEnd+0.06,\yA-0.04) arc (180:0:0.09);
  \draw[line width=0.5pt] (\xEnd+0.15,\yA-0.04) -- (\xEnd+0.23,\yA+0.07);
  \draw[line width=0.4pt] (\xEnd+0.30,\yA+0.03) -- (\xEnd+0.55,\yA+0.03);
  \draw[line width=0.4pt] (\xEnd+0.30,\yA-0.03) -- (\xEnd+0.55,\yA-0.03);
\end{scope}
\node[font=\scriptsize,mut,anchor=west] at (3.5,\pC-1.5) {Small quantum network (e.g. 45-100 parameters)};
\node[font=\small] at (5.7,\pC-1.95) {%
  \textcolor{watercol}{\faTint}\quad
  \textcolor{boltcol}{\faBolt}\quad
  \textcolor{moneycol}{\textbf{\$}}};
\draw[arr] (\ARx,\pC) -- (\ARx+\ARRLEN,\pC);
\outcome{\OUTX}{\pC}{good}{Good generalization}
 
\end{tikzpicture}%
  }
  \caption{\textbf{Learning regimes and data efficiency.}
  \emph{Top (blue):} with a large training set, a deep network with
  $\mathcal{O}(10^5)$ parameters generalizes well, but at high data and
  resource cost (water, energy, money). \emph{Middle (red):} given only a
  small dataset, the same large model overfits and generalizes poorly, and
  a smaller classical network (still thousands of parameters) fares no
  better. \emph{Bottom (green):} a quantum convolutional network with only
  e.g. 45-100 parameters learns from the same small dataset and small resource cost (water, energy, money) while
  generalizing well, illustrating the data-efficient regime studied in this
  work.}
  \label{fig:intro}
\end{figure*}

Access to high-quality labelled datasets of sufficient size is limited in many medical applications, which continues to constrain the performance of machine learning in clinical settings ~\cite{varoquaux2022machine, schafer2024overcoming}. Modern deep neural networks reach high accuracy only with large high-quality training sets, substantial compute, and extensive architectural tuning, at a high computational 
and energy cost~\cite{tripp2024measuring}. Datasets of this scale are seldom available in clinical practice, where annotated images are expensive to acquire, expert labelling is slow, privacy regulation restricts data sharing, and many conditions are rare, so the datasets available for training are small and often imbalanced. Models developed for the data-rich setting tend to overfit and generalize poorly when trained on small datasets (Figure~\ref{fig:intro}), which has motivated work on data-efficient and few-shot learning for medical imaging ~\cite{pachetti2024systematic}. The relevant question is then not how far a model's accuracy scales with additional data, but how much it can learn from few examples.

Quantum machine learning (QML) has been proposed as one route to learning from scarce data~\cite{biamonte2017quantum,cerezo2021variational}. Rather than relying on
ever-larger parameter counts, QML seeks computational and statistical leverage
from the structure of quantum states and operations. Recent reviews of QML for medical imaging frame the field less around raw accuracy than around possible
gains in data efficiency and inductive bias under data-limited
deployment~\cite{wei2023quantum,yan2024review}. Among QML architectures, quantum
convolutional neural networks (QCNNs) are especially attractive. Introduced by Cong, Choi, and Lukin~\cite{cong2019quantum}, a QCNN alternates
convolutional and pooling blocks that compress information while preserving
task-relevant features, combining parameter sharing and hierarchical pooling in a
way that parallels the layered, weight-shared structure of
classical convolutional networks (the sense in which ``locality'' applies here is
made precise in \emph{QCNN Architecture section}). Crucially, Caro
\emph{et al.}~\cite{caro2022generalization} proved that the generalization error
of a parameterized quantum model scales with the number of trainable gates rather
than with the exponentially large Hilbert-space dimension, placing QCNNs, whose
number of independent gates grows only logarithmically with system size, in a
particularly favorable, potentially sample-efficient regime. A demonstration by \citet{qml_generalization_tutorial} operationalized this
prediction on a binary digit-recognition task, providing a compelling proof of
principle for learning from very few samples.

These guarantees however, are subject to important qualifications. Gil-Fuster, Eisert, and
Bravo-Prieto~\cite{gil2024understanding} showed that uniform, complexity-style
bounds cannot by themselves explain the observed generalization of quantum models,
which can also fit randomized labels. Bermejo \emph{et
al.}~\cite{bermejo2026quantum} demonstrated that QCNNs are effectively classically
simulable on the locally easy benchmarks where they perform well, cautioning
against reading strong accuracy on simple datasets as evidence of a uniquely
quantum capability. These findings do not diminish the importance of
data-efficient learning; rather, they sharpen the question that motivates this
work: under what conditions does a realistic, hardware-compatible QCNN learn
meaningful structure from limited data, and which factors constrain the scaling of such models to higher-dimensional inputs.

We address this question empirically and make three contributions. First, we
implement a hardware-compatible QCNN as a fully dynamic Qiskit circuit with
mid-circuit measurement and classical feed-forward, consistent with IBM's
dynamic-circuit model~\cite{carrera2024combining}, and show on a binary digit task that strong test performance is attainable from as few as \(10\) training
samples, with the generalization error decreasing as the training set grows.
Second, we move beyond accuracy and quantify the dominant scaling bottleneck for
NISQ-era image classification through a detailed transpilation analysis of
amplitude and angle encoding across resolutions from \(2\times2\) to
\(512\times512\) pixels, revealing a sharp asymmetry between qubit efficiency and
circuit depth. Third, we extend the study to BreastMNIST as a medically motivated benchmark and compare the QCNN against classical convolutional networks at both matched and unconstrained parameter budgets. Our aim is explicitly not to claim quantum advantage, but to establish where the few-shot generalization story holds,
where encoding rather than optimization becomes the binding constraint, and whether a highly parameter-efficient QCNN can still learn clinically relevant signal.

The small-data regime studied here is characteristic of many problems in the life sciences.
Molecular structure prediction and design has reached today a mature development thanks to the large amount of protein sequence and structure data available. Nonetheless, not all molecular entities, especially those relevant for therapeutic applications, are so well represented as proteins, rendering AI-driven approaches much more challenging. This is the case for instance of non-canonical amino acids and nucleic acids, whose data coverage is more sparse resulting in poor structure prediction and design results ~\cite{abriata2026casp16}. Also, in the context of clinical trial optimization, where predictive models rely only on small amount of compounds or clinical results. These problems are characterized by scarce, costly-to-acquire data and modest feature dimensionality, precisely the regime in which the data efficiency and structured inductive bias of QCNNs are most likely to matter.

\section{Background\label{sec:background}}

\begin{figure*}[!t]
  \centering
  \includegraphics[width=\textwidth]{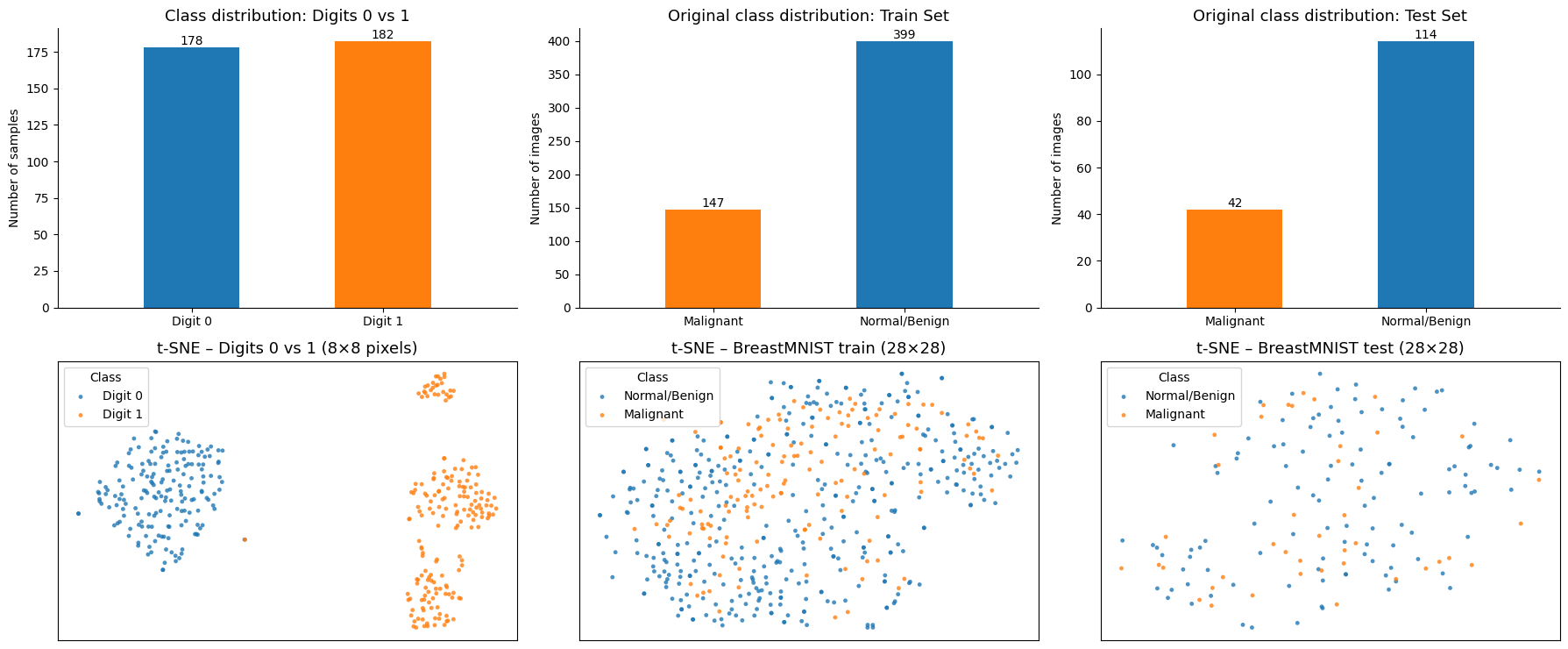}
  \caption{\textbf{Dataset composition and structure.}
  \textbf{Top:} class distributions. The digit task ($0$ vs.\ $1$) is balanced ($178$ and $182$ samples), whereas BreastMNIST is class-imbalanced in both its training ($147$ malignant, $399$ normal/benign) and test ($42$ malignant, $114$ normal/benign) splits, motivating the fixed class-balanced test set and the
  balanced-accuracy metric used for that task. \textbf{Bottom:} two-dimensional t-SNE embeddings of the flattened images at their native resolution (digits $8\times8$; BreastMNIST $28\times28$), coloured by class. The digit classes form two well-separated clusters, whereas the BreastMNIST classes
  overlap substantially in pixel space.}
  \label{fig:tsne}
\end{figure*}

\begin{figure*}[t]
    \centering
    \begin{minipage}[t]{0.48\textwidth}
        \centering
        \includegraphics[width=\linewidth]{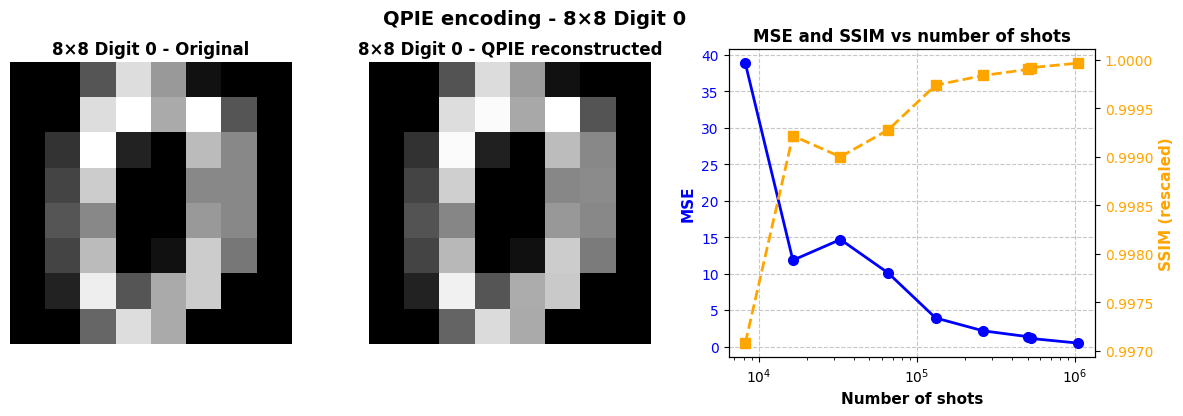}
        \vspace{0.3em}
        \textbf{(a)} QPIE encoding and reconstruction for an \(8\times8\) digit 0 sample.
    \end{minipage}
    \hfill
    \begin{minipage}[t]{0.48\textwidth}
        \centering
        \includegraphics[width=\linewidth]{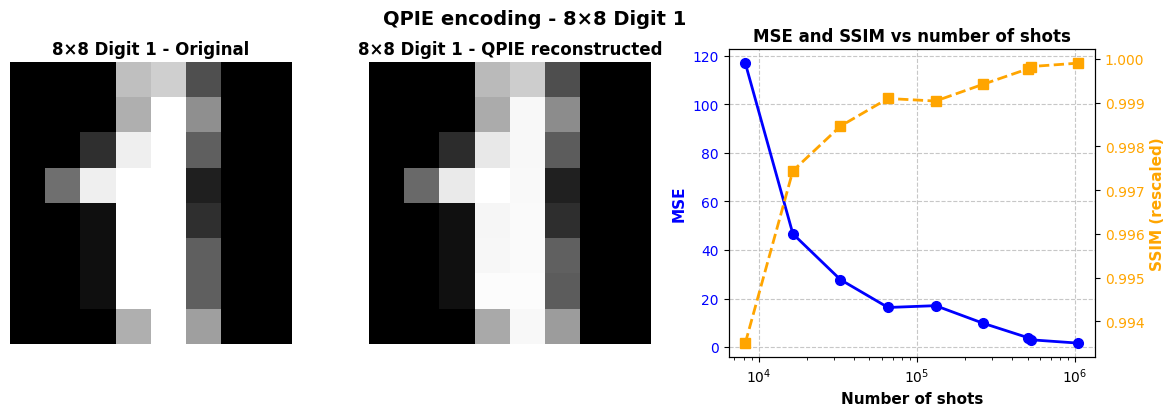}
        \vspace{0.3em}
        \textbf{(b)} QPIE encoding and reconstruction for an \(8\times8\) digit 1 sample.
    \end{minipage}

    \vspace{1em}

    \begin{minipage}[t]{0.48\textwidth}
        \centering
        \includegraphics[width=\linewidth]{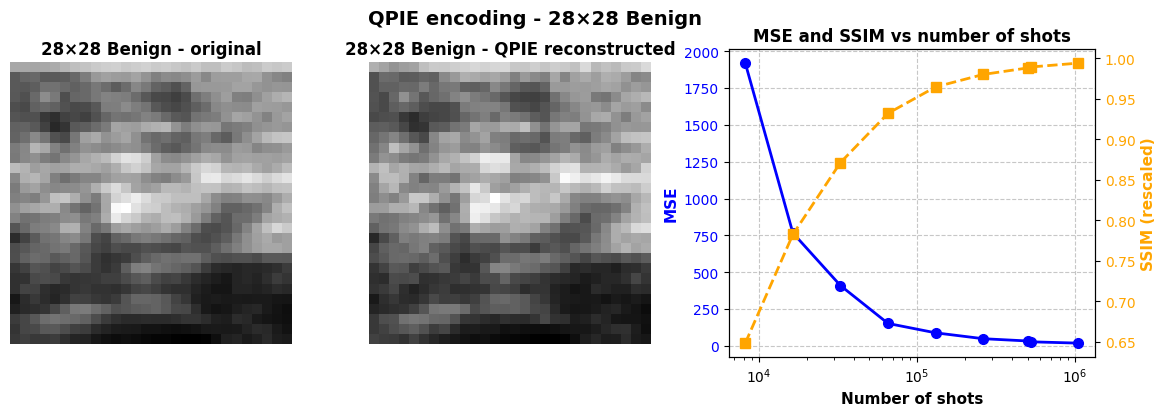}
        \vspace{0.3em}
        \textbf{(c)} QPIE encoding and reconstruction for a \(28\times28\) benign BreastMNIST sample.
    \end{minipage}
    \hfill
    \begin{minipage}[t]{0.48\textwidth}
        \centering
        \includegraphics[width=\linewidth]{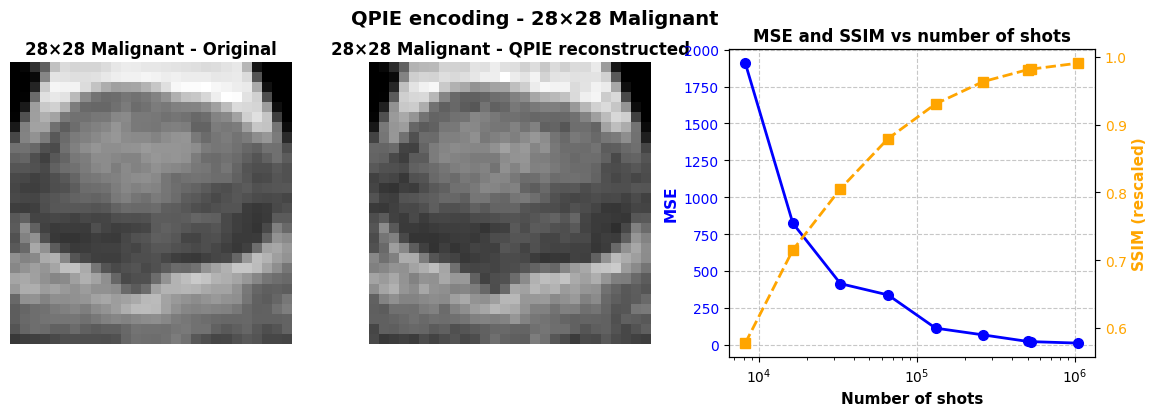}
        \vspace{0.3em}
        \textbf{(d)} QPIE encoding and reconstruction for a \(28\times28\) malignant BreastMNIST sample.
    \end{minipage}

    \caption{\footnotesize
    \textbf{Visualization of QPIE encoding fidelity across representative samples from the sklearn digits dataset and BreastMNIST.} Each panel shows the original image (left), the image reconstructed after QPIE-based encoding and measurement (center), and the corresponding reconstruction quality as a function of the number of shots (right), quantified using mean squared error (MSE) and structural similarity index measure (SSIM). For both low-resolution digit images and higher-resolution BreastMNIST samples, increasing the number of shots improves reconstruction fidelity, with MSE decreasing and SSIM increasing consistently. These results illustrate both the robustness of QPIE under finite sampling and the higher shot requirements associated with more complex medical image inputs.}
    \label{fig:qpie_examples}
\end{figure*}

\subsection{Generalization in quantum machine learning}
A trained model is useful only insofar as it performs well on data it has not
seen. For a parameterized quantum model with parameters \(\alpha\), performance on
a single labeled example \((x,y)\) is measured by a loss \(\ell(\alpha; x, y)\).
Two averages of this loss are relevant. The \emph{empirical loss}
\begin{equation}
\hat{R}_N(\alpha) = \frac{1}{N}\sum_{i=1}^{N} \ell(\alpha; x_i, y_i)
\label{eq:emp_loss}
\end{equation}
is the mean loss over the \(N\) training examples, while the \emph{expected loss}
\begin{equation}
R(\alpha) = \mathbb{E}_{(x,y)\sim P}\,\bigl[\ell(\alpha; x, y)\bigr]
\label{eq:exp_loss}
\end{equation}
is the mean loss over the underlying data distribution \(P\). Since \(P\) is
unknown, training minimizes the empirical loss \(\hat{R}_N(\alpha)\) as a proxy for
the expected loss \(R(\alpha)\); the validity of that proxy is controlled by the
\emph{generalization error}
\begin{equation}
\mathrm{gen}(\alpha) = R(\alpha) - \hat{R}_N(\alpha),
\label{eq:gen}
\end{equation}
the gap between performance on unseen data and performance on the training set. A small generalization error means that low training loss reliably implies low loss at deployment. Caro \emph{et al.}~\cite{caro2022generalization} bounded \(\mathrm{gen}(\alpha)\)
for variational quantum models through the structure of the circuit rather than
the dimension of the Hilbert space. For a model built from \(T\) independently
parameterized local gates, they proved that, with high probability over the draw
of the training set,
\begin{equation}
\mathrm{gen}(\alpha) \in \mathcal{O}\!\left(\sqrt{\frac{T \log T}{N}}\,\right),
\label{eq:caro_basic}
\end{equation}
so that the error becomes small once \(N\) grows in proportion to \(T \log T\)
(more precisely, \(N \sim T\log T/\varepsilon^{2}\) samples guarantee
\(\mathrm{gen}(\alpha)\le\varepsilon\)). When each gate is reused at most \(M\)
times, as occurs under the translational symmetry of a convolutional layer or when the circuit is executed repeatedly to estimate expectation values, the bound
depends only logarithmically on \(M\),
\begin{equation}
\mathrm{gen}(\alpha) \in \mathcal{O}\!\left(\sqrt{\frac{T \log (MT)}{N}}\,\right).
\label{eq:caro_reuse}
\end{equation}
A third bound accounts for the optimization trajectory: if only \(K \ll T\) gates change appreciably during training, the effective complexity is governed by \(K\) rather than \(T\), and the error sharpens to \(\mathcal{O}(\sqrt{K/N})\). 
Because these bounds are distribution-independent and depend only on circuit structure, they apply directly to QCNNs. A QCNN applies translationally invariant convolutional unitaries and contracts the register through pooling, so its number of independent parameters grows only as \(T \in \mathcal{O}(\log n)\) in the qubit count \(n\). Inserting this scaling into Eq.~\eqref{eq:caro_reuse} yields a polylogarithmic sample complexity, \(N \in \mathcal{O}(\log^2 n)\), for which Caro \emph{et al.} proved and confirmed numerically on a quantum-phase-recognition task, that good generalization is attainable. Two caveats govern how this transfers to the present study. First, the guarantee and its numerical demonstration concern the classification of quantum states across a phase transition; whether the same small-sample behaviour survives once \emph{classical} images are loaded into the circuit is an empirical question, which we take up in the \emph{Results}. Second, a small generalization error is informative only if the model can be trained at all, and QCNNs are known to avoid the barren plateaus that obstruct gradient-based optimization in many quantum neural networks: their gradient variance vanishes no faster than polynomially in
\(n\)~\cite{pesah2021absence}.

Other generalization frameworks exist, including margin-based bounds that relate
the error to the distribution of classification margins~\cite{hur2024understanding}
and information-theoretic bounds expressed through the (R\'enyi) mutual information between the encoded quantum states and the data or class
labels~\cite{banchi2021generalization}. We organize the present work around the
framework of Caro \emph{et al.} because it provides concrete, structure-dependent
guarantees for the parameter-reusing variational circuits that run on near-term hardware, and because its dependence on the number of trainable gates rather than the Hilbert-space dimension is precisely the property that the convolutional and pooling structure of a QCNN is designed to exploit.

\subsection{Encoding classical data into quantum states}
Applying a QCNN to classical images first requires a map from a feature vector to
a quantum state, and at scale it is this step, rather than the variational layers,
that dominates the resource budget~\cite{lang2024representation,balewski2024quantum}.

Encoding strategies differ mainly in where the classical information is placed in
the quantum state, which sets their qubit and gate
cost~\cite{lang2024representation}. In \emph{basis encoding}, a value is written
into a computational basis state, using one qubit per bit of data. In
\emph{amplitude encoding}, a normalized \(d\)-dimensional vector is written into
the amplitudes of \(n = \lceil \log_2 d \rceil\) qubits~\cite{schuld2021machine};
this is maximally qubit-efficient, so that a \(64\)-pixel image occupies only
\(6\) qubits, but a generic vector requires a state-preparation circuit of up to
\(\mathcal{O}(2^n)\) gates~\cite{shende2006synthesis}, whose depth grows steeply
with resolution. In \emph{angle encoding}, each value sets the rotation angle of a
single qubit, giving a constant-depth circuit at the cost of one qubit per value.
Several image-specific representations build on these primitives: the flexible
representation of quantum images (FRQI) stores pixel intensities in rotation angles
together with position qubits, the novel enhanced quantum representation (NEQR)
stores intensities in a separate basis-encoded register, and quantum probability
image encoding (QPIE) stores normalized intensities \(c_k = f_k/\lVert f\rVert_2\)
directly in the amplitudes of \(\lceil \log_2 P \rceil\) qubits for a \(P\)-pixel
image~\cite{yao2017quantum,lang2024representation}. QPIE is thus nothing other than
amplitude encoding applied to an image: the two specify the same state-preparation
map, and in the context of image encoding we treat ``QPIE'' and ``amplitude
encoding'' as synonyms throughout this work.

Concretely, each image is flattened in row-major order to a vector
$f\in\mathbb{R}^{P}$ (zero-padded to the next power of two when $P$ is not one), and the
normalized intensity $c_k=f_k/\lVert f\rVert_2$ of the pixel at flattened index $k$ is
stored in the amplitude of basis state $\lvert k\rangle$, giving
$\lvert\psi\rangle=\sum_{k}c_k\lvert k\rangle$. This map is a lossless bijection: the
pixel position is fully recoverable from the index, with $\mathrm{row}=\lfloor k/W\rfloor$
and $\mathrm{col}=k \bmod W$ for image width $W$. Writing $k=\sum_j b_j 2^j$, qubit $j$
carries a single address bit $b_j$ of the flattened index, so one qubit corresponds to one
binary digit of the pixel address rather than to a pixel or a contiguous patch.

For all classification experiments in this work we use amplitude encoding (QPIE), which keeps the qubit count logarithmic in the number of pixels. To map the broader design space, however, our scaling analysis spans both encodings. The two ends of this spectrum, narrow (i.e. qubit-frugal) but deep versus shallow but wide (i.e. qubit-hungry), fix the hardware envelope within which a QCNN pipeline can run at a given resolution. We map that envelope quantitatively across encodings and image sizes in the \emph{Scaling behavior} section.

\section{Methods\label{sec:methods}}
\subsection{Dataset and preprocessing}

We evaluated the QCNN on two binary image-classification tasks of differing resolution and difficulty, which were implemented on different backends.

The primary benchmark was the scikit-learn \texttt{load\_digits}
dataset~\cite{scikit-learn}, which contains $8\times8$ grayscale images of handwritten digits; we retained the two classes $0$ and $1$ ($178$ and $182$ samples, respectively). Each image was flattened to a $64$-dimensional vector and
$\ell_2$-normalized, then amplitude-encoded into a $6$-qubit register ($2^{6}=64$). Training sets of sizes $N\in\{2,5,10,20,40,80\}$ were drawn by stratified sampling so that the two classes were balanced. For each repetition, a separate test set of $100$ samples was drawn at random from the remaining images; training and test sets were thus disjoint within a repetition, but the test set was re-sampled across
repetitions rather than held fixed. Metrics are reported as means over $10$ independent repetitions per training size, each using a fresh data split and randomized weight initialization. This task was implemented in Qiskit as a dynamic circuit with mid-circuit measurement and classical feed-forward.

To probe generalization in a clinically relevant domain, we extended the study to
BreastMNIST from the MedMNIST~v2 collection~\cite{medmnist}, which comprises
$28\times28$ grayscale breast-ultrasound images under the standard binary task
that distinguishes malignant from normal and benign images. In contrast to the
digits experiment, BreastMNIST was processed at its native $28\times28$
resolution: each image was flattened to a $784$-dimensional vector, zero-padded to
$1024$ ($2^{10}$) and $\ell_2$-normalized, then amplitude-encoded into a
$10$-qubit register. Training sets of sizes $N\in\{20,40,80,160,294\}$ were drawn
by stratified sampling, and models were evaluated on a fixed, class-balanced test
set of $84$ samples, so that accuracy equals balanced accuracy, averaged over
$20$ independent repetitions per training size. Figure~\ref{fig:tsne} shows the class composition of both datasets together with a
two-dimensional t-SNE embedding of the raw pixel vectors: the two digit classes form
clearly separated clusters, whereas the malignant and normal/benign breast-ultrasound
classes overlap substantially in pixel space, indicating a more difficult
classification task in the clinical domain. 

A complementary axis, beyond the scope of this work, is the expressivity of the encoding rather than only its cost. Because the encoding fixes the model's accessible frequency spectrum~\cite{schuld2021effect}, a fixed, qubit-frugal amplitude encoding may supply the frequencies needed for the cleanly separable digit task more readily than for the overlapping BreastMNIST classes, whose decision boundary plausibly demands higher frequencies, independently of parameter count. Encodings that deliberately enlarge this
spectrum, such as quantum-Fourier-transform-based or Hamiltonian schemes, are therefore a natural route to greater expressivity on harder tasks, with the caveat that many such Fourier-series models are efficiently approximable
classically~\cite{landman2022classically,schreiber2023classical}.

Because the $10$-qubit state
preparation makes an equivalent Qiskit dynamic circuit substantially more
resource-intensive, this task was implemented in PennyLane using statevector
simulation.
Figure~\ref{fig:qpie_examples} visualizes a representative sample from each class
of the two datasets used in our experiments, the handwritten digits $0$ and $1$
and a benign and a malignant breast-ultrasound image, alongside their amplitude
(QPIE) encoding and reconstruction.

\subsection{Data encoding}

Amplitude encoding was used to prepare classical image data as quantum 
states~\cite{schuld2021machine}. The $\ell_2$-normalized feature vector 
$\tilde{\mathbf{x}} \in \mathbb{R}^{2^n}$ is mapped onto the computational basis 
amplitudes of an $n$-qubit register:
\begin{equation}
    |\psi\rangle = \sum_{i=0}^{2^n - 1} \tilde{x}_i \, |i\rangle, 
    \qquad \sum_i |\tilde{x}_i|^2 = 1.
    \label{eq:amplitude_encoding}
\end{equation}
This encoding is maximally qubit-efficient: a 64-pixel image requires only 
$n = 6$ qubits. The principal cost is in state preparation, which requires 
$\mathcal{O}(2^n)$ gates in the worst case~\cite{shende2006synthesis}, a 
bottleneck that grows exponentially with image resolution and motivates the 
hardware feasibility analysis discussed in the \emph{Scaling behavior} section. 
The encoding was implemented via Qiskit's \texttt{initialize()} instruction, which is functionally equivalent to PennyLane's \texttt{AmplitudeEmbedding}.

\subsection{QCNN Architecture}

\begin{figure*}[t]
  \centering
  \resizebox{\linewidth}{!}{%
  \begin{quantikz}[column sep=7pt, row sep=14pt]
    \lstick{$q_0$} & \gate[6]{\usebox{\embedbox}}
      & \gate[style={fill=convcol}]{U(\theta_0,\theta_1,\theta_2)}
      & \gate[2,style={fill=convcol}]{R_{XX}(\theta_6)} & \gate[2,style={fill=convcol}]{R_{YY}(\theta_7)} & \gate[2,style={fill=convcol}]{R_{ZZ}(\theta_8)}
      & \gate[style={fill=convcol}]{U(\theta_9,\theta_{10},\theta_{11})}
      & \qw & \qw & \qw & \qw
      & \gate[style={fill=poolcol}]{U(\theta_{15},\theta_{16},\theta_{17})}\vcw{1} & \qw
      & \gate[3,style={fill=convcol}]{R_{XX}(\theta_{18})} & \gate[3,style={fill=convcol}]{R_{YY}(\theta_{19})} & \gate[3,style={fill=convcol}]{R_{ZZ}(\theta_{20})}
      & \gate[style={fill=convcol}]{U(\theta_{21},\theta_{22},\theta_{23})}
      & \qw & \qw & \qw & \qw
      & \gate[style={fill=poolcol}]{U(\theta_{27},\theta_{28},\theta_{29})}\vcw{2} & \qw
      & \gate[5,style={fill=densecol}]{U_{\mathrm{SU(4)}}\,(\theta_{30}\text{--}\theta_{44})} & \meter{} & \setwiretype{c} & \rstick{$\begin{array}{l}0 \rightarrow \text{Outcome 1}\\[2pt] 1 \rightarrow \text{Outcome 2}\end{array}$}\qw \\
    \lstick{$q_1$} &
      & \gate[style={fill=convcol}]{U(\theta_3,\theta_4,\theta_5)}
      & & &
      & \gate[style={fill=convcol}]{U(\theta_{12},\theta_{13},\theta_{14})}
      & \gate[2,style={fill=convcol}]{R_{XX}(\theta_6)} & \gate[2,style={fill=convcol}]{R_{YY}(\theta_7)} & \gate[2,style={fill=convcol}]{R_{ZZ}(\theta_8)}
      & \gate[style={fill=convcol}]{U(\theta_9,\theta_{10},\theta_{11})}
      & \meter[style={fill=poolcol}]{} & \setwiretype{c} \\
    \lstick{$q_2$} &
      & \gate[style={fill=convcol}]{U(\theta_0,\theta_1,\theta_2)}
      & \gate[2,style={fill=convcol}]{R_{XX}(\theta_6)} & \gate[2,style={fill=convcol}]{R_{YY}(\theta_7)} & \gate[2,style={fill=convcol}]{R_{ZZ}(\theta_8)}
      & \gate[style={fill=convcol}]{U(\theta_9,\theta_{10},\theta_{11})}
      & & & & \gate[style={fill=convcol}]{U(\theta_{12},\theta_{13},\theta_{14})}
      & \gate[style={fill=poolcol}]{U(\theta_{15},\theta_{16},\theta_{17})}\vcw{1} & \qw
      & & & & \gate[style={fill=convcol}]{U(\theta_{24},\theta_{25},\theta_{26})}
      & \gate[3,style={fill=convcol}]{R_{XX}(\theta_{18})} & \gate[3,style={fill=convcol}]{R_{YY}(\theta_{19})} & \gate[3,style={fill=convcol}]{R_{ZZ}(\theta_{20})}
      & \gate[style={fill=convcol}]{U(\theta_{21},\theta_{22},\theta_{23})}
      & \meter[style={fill=poolcol}]{} & \setwiretype{c} \\
    \lstick{$q_3$} &
      & \gate[style={fill=convcol}]{U(\theta_3,\theta_4,\theta_5)}
      & & &
      & \gate[style={fill=convcol}]{U(\theta_{12},\theta_{13},\theta_{14})}
      & \gate[2,style={fill=convcol}]{R_{XX}(\theta_6)} & \gate[2,style={fill=convcol}]{R_{YY}(\theta_7)} & \gate[2,style={fill=convcol}]{R_{ZZ}(\theta_8)}
      & \gate[style={fill=convcol}]{U(\theta_9,\theta_{10},\theta_{11})}
      & \meter[style={fill=poolcol}]{} & \setwiretype{c} \\
    \lstick{$q_4$} &
      & \gate[style={fill=convcol}]{U(\theta_0,\theta_1,\theta_2)}
      & \gate[2,style={fill=convcol}]{R_{XX}(\theta_6)} & \gate[2,style={fill=convcol}]{R_{YY}(\theta_7)} & \gate[2,style={fill=convcol}]{R_{ZZ}(\theta_8)}
      & \gate[style={fill=convcol}]{U(\theta_9,\theta_{10},\theta_{11})}
      & & & & \gate[style={fill=convcol}]{U(\theta_{12},\theta_{13},\theta_{14})}
      & \gate[style={fill=poolcol}]{U(\theta_{15},\theta_{16},\theta_{17})}\vcw{1} & \qw
      & \qw & \qw & \qw & \qw
      & & & & \gate[style={fill=convcol}]{U(\theta_{24},\theta_{25},\theta_{26})}
      & \qw & \qw & \qw \\
    \lstick{$q_5$} &
      & \gate[style={fill=convcol}]{U(\theta_3,\theta_4,\theta_5)}
      & & &
      & \gate[style={fill=convcol}]{U(\theta_{12},\theta_{13},\theta_{14})}
      & \qw & \qw & \qw & \qw
      & \meter[style={fill=poolcol}]{} & \setwiretype{c}
  \end{quantikz}%
  }
 
  \vspace{6pt}
  \begin{tikzpicture}[font=\footnotesize, every node/.style={inner sep=2pt}]
    \node[fill=convcol, draw, minimum width=0.5cm, minimum height=0.3cm] (s1) {};
    \node[right=2pt of s1] (l1) {Convolution layer};
    \node[fill=poolcol, draw, minimum width=0.5cm, minimum height=0.3cm, right=16pt of l1] (s2) {};
    \node[right=2pt of s2] (l2) {Pooling layer (mid-circuit measurement)};
    \node[fill=densecol, draw, minimum width=0.5cm, minimum height=0.3cm, right=16pt of l2] (s3) {};
    \node[right=2pt of s3] (l3) {Dense $SU(4)$ layer};
  \end{tikzpicture}
 
  \caption{\textbf{QCNN architecture implemented in Qiskit for \(8 \times 8\) grayscale inputs.} 
  An $8\times8$ input image (grey) is amplitude-embedded into 6 qubits, then processed by
  a brick-wall convolution layer (blue), measurement-based pooling (green), a second
  convolution and pooling stage, a dense $SU(4)$ block (yellow), and a final readout of
  $q_0$ into a classical register encoding the outcome class. Double lines denote
  classical control; identical gate labels denote shared parameters. The model has
  $15+3+9+3+15 = 45$ trainable parameters.}
  \label{fig:circuit}
\end{figure*}
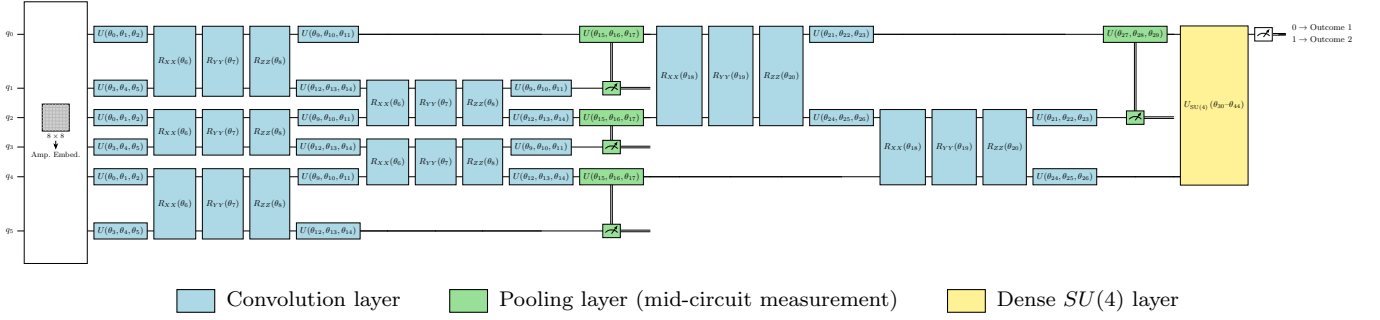

The QCNN follows the hierarchical design of Cong, Choi, and 
Lukin~\cite{cong2019quantum}, with alternating convolutional and pooling layers 
followed by a dense layer (Figure~\ref{fig:circuit}). For $8 \times 8$ input images 
the circuit operates on $n = 6$ qubits, with 51 gates, 45 parameters and a logical circuit depth of 23 (before transpilation). Because the convolutional and pooling blocks are defined by their action on nearest-neighbour qubit pairs rather than by a fixed register size, the same construction applies unchanged to larger images: the qubit count grows only as $\lceil \log_2 P \rceil$ with the pixel count $P$, while weight sharing keeps the number of trainable parameters within each layer fixed, so the architecture extends to higher-resolution inputs without modification.

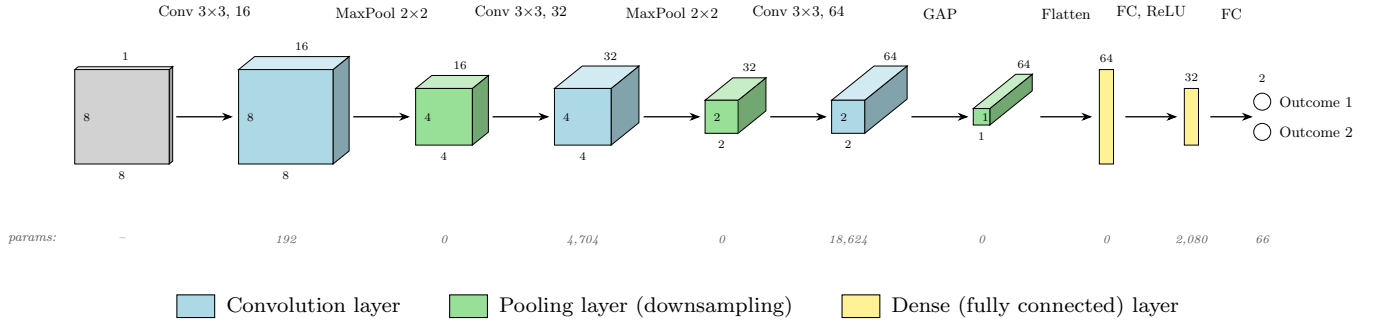
\begin{figure*}[t]
\centering
\resizebox{\linewidth}{!}{%
\begin{tikzpicture}[font=\sffamily]
 
\fmap{0.0}{0}{2.0}{0.12}{inputcol}{8}{8}{1}{--}
\fmap{3.47}{0}{2.0}{0.55}{convcol}{8}{8}{16}{192}
\fmap{7.22}{0}{1.2}{0.55}{poolcol}{4}{4}{16}{0}
\fmap{10.16}{0}{1.2}{0.95}{convcol}{4}{4}{32}{4{,}704}
\fmap{13.35}{0}{0.7}{0.95}{poolcol}{2}{2}{32}{0}
\fmap{16.03}{0}{0.7}{1.45}{convcol}{2}{2}{64}{18{,}624}
\fmap{19.03}{0}{0.35}{1.45}{poolcol}{1}{1}{64}{0}
 
\draw[-{Stealth},thick] (2.15,0) -- (3.35,0);
\draw[-{Stealth},thick] (5.90,0) -- (7.10,0);
\draw[-{Stealth},thick] (8.85,0) -- (10.05,0);
\draw[-{Stealth},thick] (12.05,0) -- (13.25,0);
\draw[-{Stealth},thick] (14.73,0) -- (15.93,0);
\draw[-{Stealth},thick] (17.72,0) -- (18.93,0);
 
\oplabel{2.75}{Conv $3{\times}3$, 16}
\oplabel{6.50}{MaxPool $2{\times}2$}
\oplabel{9.45}{Conv $3{\times}3$, 32}
\oplabel{12.65}{MaxPool $2{\times}2$}
\oplabel{15.33}{Conv $3{\times}3$, 64}
\oplabel{18.33}{GAP}
 
\draw[-{Stealth},thick] (20.45,0) -- (21.55,0);
\oplabel{21.0}{Flatten}
 
\def\fcx{21.7}
\fill[densecol,draw=black,line width=0.3pt] (\fcx,-1.0) rectangle (\fcx+0.30,1.0);
\node[anchor=south, font=\scriptsize] at (\fcx+0.15,1.05) {64};
\node[anchor=north, font=\scriptsize\itshape, paramcol] at (\fcx+0.15,\parbase) {0};
\draw[-{Stealth},thick] (\fcx+0.55,0) -- (\fcx+1.65,0);
\oplabel{\fcx+1.1}{FC, ReLU}
\fill[densecol,draw=black,line width=0.3pt] (\fcx+1.8,-0.6) rectangle (\fcx+2.1,0.6);
\node[anchor=south, font=\scriptsize] at (\fcx+1.95,0.65) {32};
\node[anchor=north, font=\scriptsize\itshape, paramcol] at (\fcx+1.95,\parbase) {2{,}080};
\draw[-{Stealth},thick] (\fcx+2.35,0) -- (\fcx+3.25,0);
\oplabel{\fcx+2.8}{FC}
\fill[white,draw=black,line width=0.6pt] (\fcx+3.45,0.32) circle (0.18);
\fill[white,draw=black,line width=0.6pt] (\fcx+3.45,-0.32) circle (0.18);
\node[anchor=south, font=\scriptsize] at (\fcx+3.45,0.62) {2};
\node[right, font=\small] at (\fcx+3.7,0.32) {Outcome 1};
\node[right, font=\small] at (\fcx+3.7,-0.32) {Outcome 2};
\node[anchor=north, font=\scriptsize\itshape, paramcol] at (\fcx+3.45,\parbase) {66};
 
\node[anchor=north, font=\footnotesize\itshape, paramcol] at (-0.9,\parbase) {params:};
 
\end{tikzpicture}%
}
 
\vspace{4pt}
\begin{center}
\begin{tikzpicture}[font=\footnotesize, every node/.style={inner sep=2pt}]
  \node[fill=convcol, draw, minimum width=0.5cm, minimum height=0.3cm] (s1) {};
  \node[right=2pt of s1] (l1) {Convolution layer};
  \node[fill=poolcol, draw, minimum width=0.5cm, minimum height=0.3cm, right=16pt of l1] (s2) {};
  \node[right=2pt of s2] (l2) {Pooling layer (downsampling)};
  \node[fill=densecol, draw, minimum width=0.5cm, minimum height=0.3cm, right=16pt of l2] (s3) {};
  \node[right=2pt of s3] (l3) {Dense (fully connected) layer};
\end{tikzpicture}
\end{center}
 
\caption{\textbf{Classical CNN baseline (8$\times$8 input).} Architecture matched to an $8\times8$ image, for visual comparison with the QCNN
of Fig.~\ref{fig:circuit}. Box dimensions are height\,$\times$\,width\,$\times$\,channels;
each $3\times3$ convolution (blue) preserves the spatial size while increasing depth
($1\!\to\!16\!\to\!32\!\to\!64$), each $2\times2$ max-pool (green) halves the spatial size
($8\!\to\!4\!\to\!2$), and the global-average-pool collapses the final $2\times2$ map to a
$64$-vector. A dense head (yellow) maps $64\!\to\!32\!\to\!2$ for the benign/malignant
decision. The spatial reduction $8\!\to\!4\!\to\!2\!\to\!1$ and weight-shared local filters
mirror the QCNN's qubit reduction $6\!\to\!3\!\to\!2$ and its parameter-shared two-qubit
convolutions; max-pooling plays the role of the QCNN's measurement-based pooling and the
fully connected head corresponds to the dense $SU(4)$ block. Italic grey numbers are
per-stage trainable parameters (convolution counts include GroupNorm affine parameters;
pooling is parameter-free), totalling $192+4{,}704+18{,}624+2{,}080+66 = 25{,}666$ --
compared with only $45$ for the QCNN.}
\label{fig:cnn}
\end{figure*}

For comparison, we benchmark the QCNN against a classical convolutional neural 
network with an analogous hierarchical structure of convolution, pooling, and a dense head (Figure~\ref{fig:cnn}); matched to an $8\times8$ input, this baseline uses $25{,}666$ trainable parameters, against $45$ for the QCNN.

\subsubsection{Convolutional layer}

The convolutional layer applies a translationally invariant two-qubit kernel 
across all nearest-neighbor qubit pairs in two alternating sweeps (even and odd), 
so that each sweep consists entirely of disjoint pairs that execute in parallel, 
keeping circuit depth constant regardless of register width. Each pair undergoes 
optional single-qubit $U_3$ pre-rotations (first layer only), a two-body 
entangling block $R_{XX}(\theta_{xx})\,R_{YY}(\theta_{yy})\,R_{ZZ}(\theta_{zz})$, 
and single-qubit $U_3$ post-rotations. All pairs share the same gate angles, so 
the layer has at most 15 independent parameters in the first layer and 9 in 
subsequent layers, irrespective of register size. This parameter sharing is the 
architectural feature that places the QCNN within the sample-efficient regime 
of Caro et al.~\cite{caro2022generalization}.

\subsubsection{Pooling layer}

The pooling layer performs measurement-based qubit reduction using mid-circuit 
measurements with classical feedback, consistent with IBM's dynamic circuit 
model~\cite{carrera2024combining}. For each qubit pair $(q_{i-1}, q_i)$, the 
qubit $q_i$ is measured and discarded; a classically conditioned $U_3$ rotation 
is applied to $q_{i-1}$ if the measurement outcome is $|1\rangle$. Three angles 
$(\theta, \phi, \lambda)$, shared across all pairs, constitute the entire 
parameter budget of this layer. Two pooling stages reduce the 6-qubit register 
to 2 active qubits.

\medskip

It is worth noting that the spatial content of the image is preserved and remains
addressable under this encoding (\emph{Encoding classical data into quantum states} subsection), but it is organized by the bit-significance scale of the flattened index rather than by a fixed spatial neighbourhood; in particular, the translational invariance of the kernel above is an invariance over the qubit register, not over the image plane. A two-qubit gate on adjacent qubits acts within a
dyadic block of the index, coupling amplitudes that differ at neighbouring bit-significance levels, and pooling removes one address bit through mid-circuit measurement and feed-forward, coarse-graining the encoded signal by one scale. Convolution and pooling thus realize a learned, dyadic multiresolution coarse-graining of the amplitude-encoded signal, in the spirit of a Haar-like hierarchy, rather than the fixed local receptive fields of a classical CNN. For power-of-two image dimensions these scales separate into row and column indices and acquire a direct spatial reading; for general dimensions the hierarchy is defined on the flattened index. Accordingly, the inductive bias it supplies, and the property rewarded by
the bound of Caro et al.~\cite{caro2022generalization}, is parameter sharing together with this hierarchical multiscale structure rather than two-dimensional spatial translation-invariance.

\subsubsection{Dense layer}

The two remaining qubits are processed by a single fully parameterized 
$\mathrm{SU}(4)$ unitary, decomposed into 15 real angles via PennyLane's 
\texttt{ArbitraryUnitary} and appended to the Qiskit circuit as a \texttt{UnitaryGate}. The total trainable parameter count is $T = 45$ for the $6$-qubit ($8\times8$) circuit. Weight sharing makes the parameter count of each layer independent of the register width, but the number of convolution and pooling stages grows with the qubit count $n$, since each stage roughly halves the number of active qubits. The total therefore increases slowly with resolution: every additional stage contributes $12$ parameters ($9$ convolutional and $3$ pooling), so the $10$-qubit ($28\times28$) circuit used for BreastMNIST has $T = 57$. This logarithmic growth, $T \in \mathcal{O}(\log n)$, is the regime in which Caro \emph{et al.}\ establish that $\mathcal{O}(\log^2 n)$ training samples suffice for bounded generalization error~\cite{caro2022generalization}.

\subsection{Training procedure}

Both experiments minimized the binary cross-entropy loss
\begin{equation}
    \mathcal{L} = -\frac{1}{M} \sum_{i=1}^{M} 
    \left[ y_i \log p^{(i)} + (1 - y_i) \log(1 - p^{(i)}) \right],
    \label{eq:loss}
\end{equation}
where $p^{(i)}$ is the probability the circuit assigns to class~$1$ for sample~$i$ and $M$ is the number of training samples, with class predictions obtained by thresholding at $0.5$. Both used the Adam optimizer~\cite{kingma2014adam} with a cosine-decayed learning-rate schedule. The two settings differ in how the circuit output and its gradients were obtained, reflecting their different backends.

\emph{Digits (Qiskit).} The QCNN was executed as a genuine dynamic circuit on IBM's \texttt{FakeKyiv} noise-aware simulator, a calibrated model of the $127$-qubit IBM~Kyiv processor, so that gate noise and finite-sampling effects were included. The output qubit $q_0$ was measured over $4{,}000$ shots to estimate 
$p = P(|1\rangle)$. Because the mid-circuit measurements are physically performed, they break the unitary continuity assumed by the parameter-shift rule~\cite{mitarai2018quantum}; gradients were therefore computed by centered finite differences with step size $\varepsilon = 0.05$. Training used an initial learning rate of $0.1$ over $50$ epochs.

\emph{BreastMNIST (PennyLane).} The $10$-qubit QCNN was instead simulated by exact statevector evaluation, which returns analytic class probabilities without sampling or device noise. The pooling measurements were realized through the deferred-measurement construction, so the circuit remained differentiable end to 
end and gradients were obtained by automatic differentiation rather than by finite differences. Training used an initial learning rate of $0.01$ over $100$ epochs. This idealized backend was chosen because the $10$-qubit amplitude-encoding state preparation makes the equivalent shot-based dynamic circuit considerably more expensive to simulate.

\emph{Statistical evaluation:} We estimate the expected loss
$R(\alpha)$ by the mean binary cross-entropy on the held-out test set and the
empirical loss $\hat{R}_N(\alpha)$ by the mean loss on the training set, so that
the per-repetition generalization gap is
$g_r = \hat{R}^{\,\mathrm{test}}_r - \hat{R}^{\,\mathrm{train}}_r$, evaluated at the
final epoch. Each quantity (loss, gap, accuracy, and balanced accuracy) is computed
independently for every repetition ($R=10$ for digits, $R=20$ for BreastMNIST) and
summarized by its mean and a $95\%$ confidence interval
$\bar{x}\pm t_{0.975,\,R-1}\,s/\sqrt{R}$, with $s$ the sample standard deviation
across repetitions and $t_{0.975,R-1}$ the Student-$t$ critical value ($2.26$ for
$R=10$, $2.09$ for $R=20$). Because the QCNN and the classical baselines are trained
and evaluated on the same per-repetition splits, all model comparisons are paired.
For the digit task the test set is re-drawn each repetition, so this interval
reflects both training- and test-set sampling; for BreastMNIST the class-balanced
$84$-image test set is held fixed across repetitions, so the interval reflects
variability from training-set resampling and weight initialization at a fixed test
set.

\emph{Backends and comparability:} The two tasks are simulated on
different backends for a computational reason quantified in \emph{Scaling} section:
the $10$-qubit amplitude-encoding state preparation required by the $28\times28$
BreastMNIST inputs transpiles to a circuit whose depth makes the shot-based, noise-calibrated dynamic-circuit simulation used for the $6$-qubit digit task prohibitively expensive, so BreastMNIST is evaluated on an idealized statevector backend. This difference does not confound our conclusions, which are drawn \emph{within} each dataset and backend and never by comparing loss or accuracy values across the two. The two settings probe complementary axes: the digit task shows that
few-shot generalization survives realistic shot noise, device noise, and
finite-difference gradients, a stringent, hardware-like test, while the BreastMNIST
task isolates the learning question on a harder, clinically motivated dataset under noise-free conditions. Because the idealized backend omits device noise, the BreastMNIST accuracies should be read as an upper estimate for that task; a
noise-calibrated or real-hardware BreastMNIST run is correspondingly identified as
future work (\emph{Discussion and conclusion} section).

\section{Results}\label{sec:results}

\begin{figure*}[!t]
  \centering
  \includegraphics[width=\textwidth]{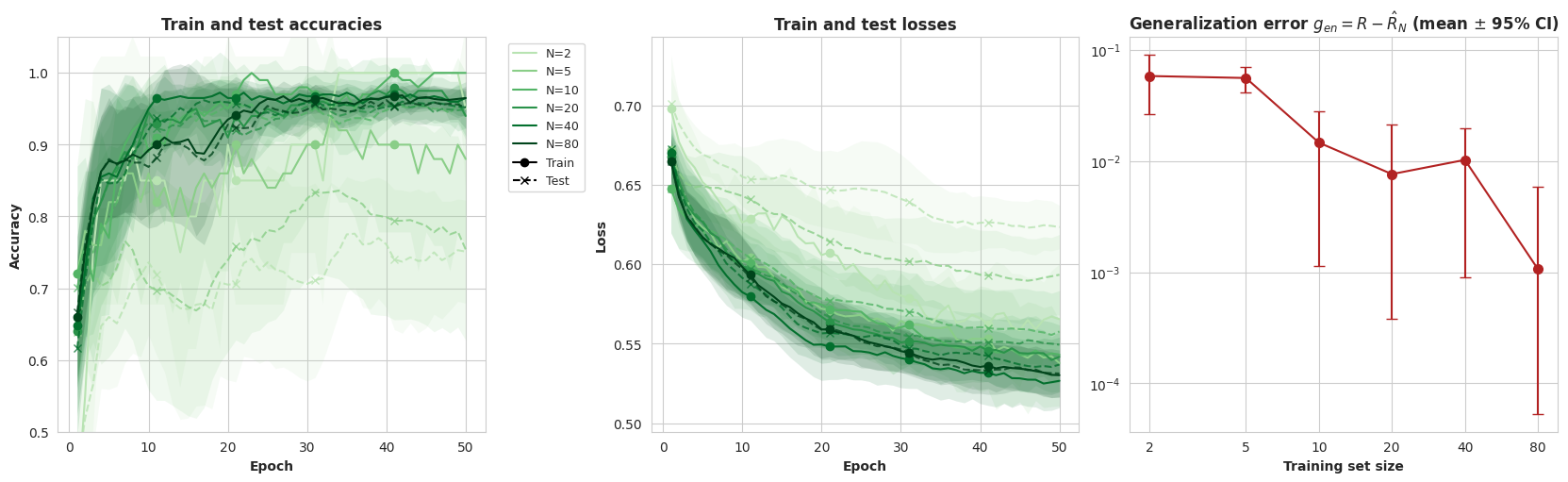}
  \caption{\textbf{Empirical generalization behavior of the QCNN across training set sizes for digits dataset.}
  \textbf{Left:} Training and test accuracy curves. While training accuracy saturates quickly, test accuracy improves with increasing training set size, indicating reduced overfitting in the larger-data regime. 
  \textbf{Center:} Training and test loss as a function of epoch for different training set sizes \(N \in \{2,5,10,20,40,80\}\). Larger training sets lead to consistently lower test loss and improved convergence.
  \textbf{Right:} Generalization error \( \mathrm{gen}(\alpha) = R(\alpha) - \hat{R}_N(\alpha) \) as a function of training set size, showing a clear decrease with increasing \(N\).
  }
  \label{fig:digits_result}
\end{figure*}

\begin{figure*}[!t]
  \centering
  \includegraphics[width=\textwidth]{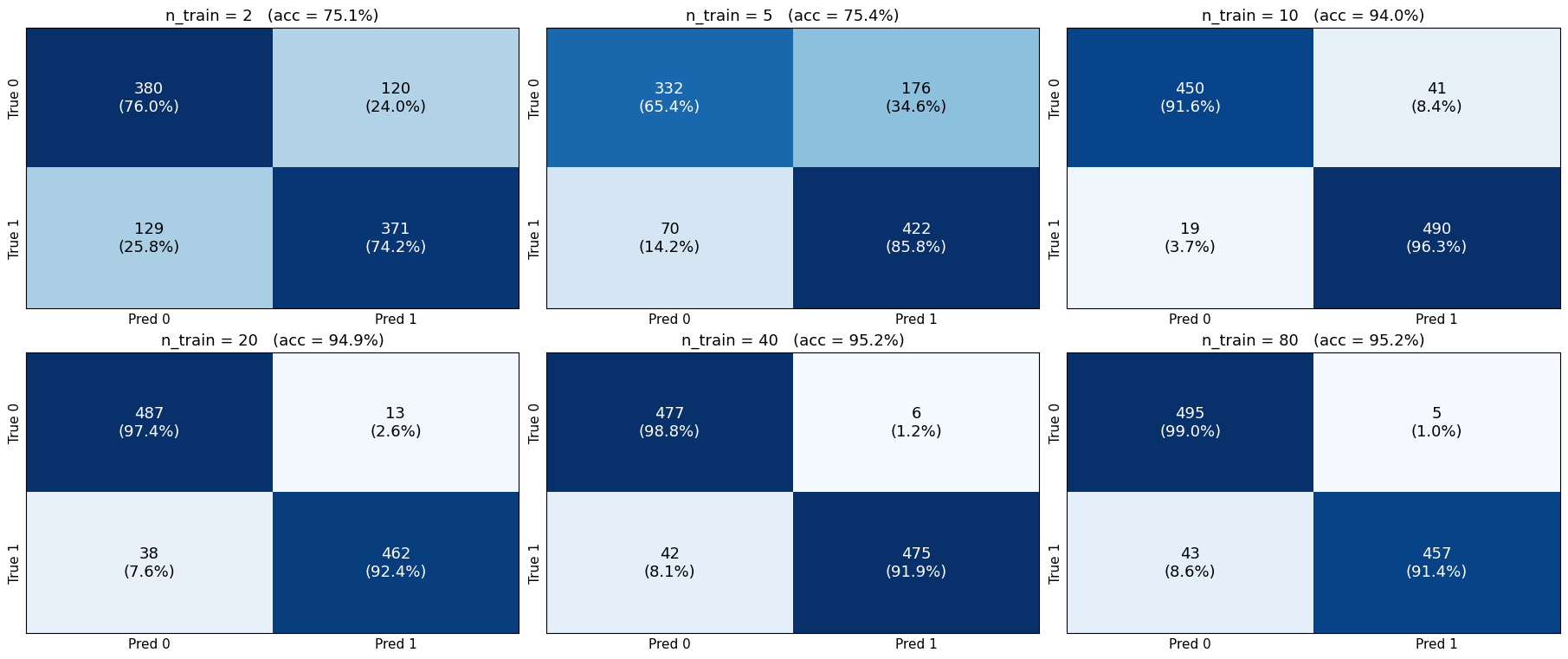}
  \caption{\textbf{Confusion matrices for the digits task across training-set sizes.}
  Aggregate counts over 10 repetitions (1000 test predictions per panel). Cell
  percentages are row-normalized, so the diagonal entries give the per-class recall.
  The direction of the error asymmetry depends on training size. At the smallest sizes it is weak or reversed (balanced at $N = 2$, class-0 misclassifications dominating at $N = 5$ and $10$), but from $N = 20$ onward, once accuracy has plateaued near 95\%, residual errors are consistently dominated by true class-1 samples predicted as class 0: class-1 recall settles around 91 to 92\% while class-0 recall exceeds 97\%.}
  \label{fig:cm_digits}
\end{figure*}

\begin{figure*}[!t]
  \centering
  \includegraphics[width=\textwidth]{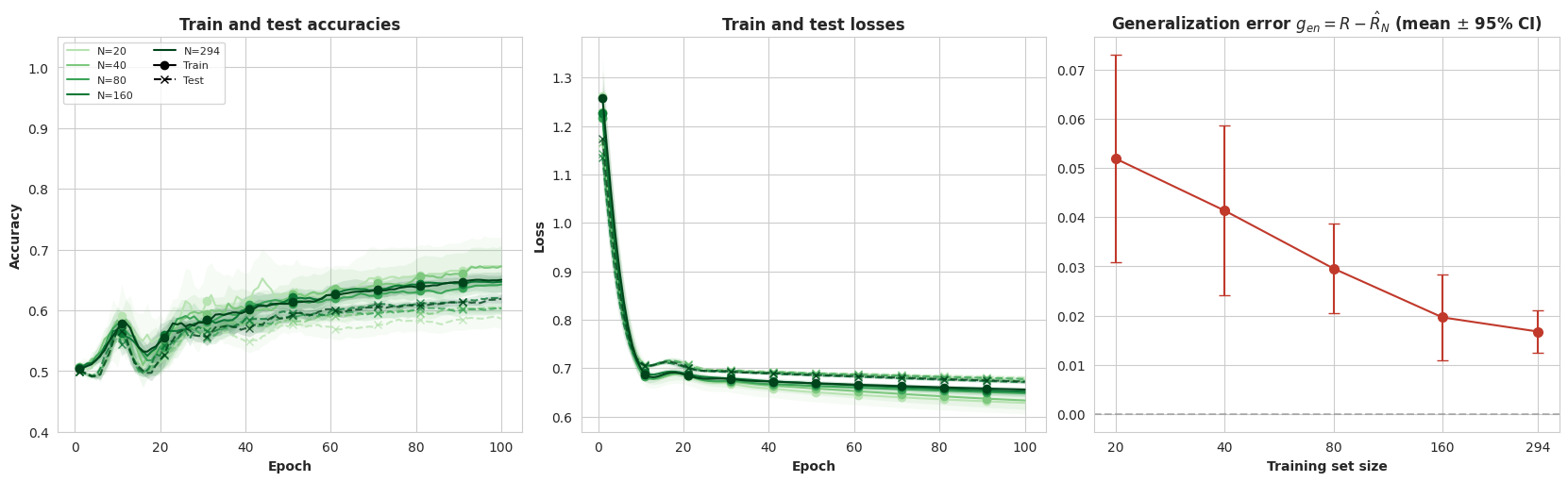}
  \caption{\textbf{Empirical generalization behavior of the QCNN across training set sizes for the BreastMNIST dataset.}
  Curves are averaged over \textbf{20} independent repetitions, with each model trained for \textbf{100} epochs.
  \textbf{Left:} Training and test accuracy curves. Training accuracy rises and stays above test accuracy, while test accuracy improves with increasing training set size and the train--test gap narrows, indicating reduced overfitting in the larger-data regime.
  \textbf{Center:} Training and test loss as a function of epoch for different training set sizes \(N \in \{20,40,80,160,294\}\). Larger training sets lead to consistently lower test loss and improved convergence.
  \textbf{Right:} Generalization error \( \mathrm{gen}(\alpha) = R(\alpha) - \hat{R}_N(\alpha) \) as a function of training set size, decreasing monotonically with increasing \(N\).
  Unlike the digits experiment (simulated in Qiskit with dynamic mid-circuit measurements), the BreastMNIST QCNN is simulated in PennyLane: the \(10\)-qubit amplitude-encoding state preparation required for the \(28\times28\) inputs makes the equivalent Qiskit dynamic-circuit implementation substantially more resource-intensive.
  }
  \label{fig:breastmnist_result}
\end{figure*}

\begin{figure*}[t]
  \centering
  \begin{subfigure}[t]{0.49\textwidth}
    \centering
    \includegraphics[width=\linewidth]{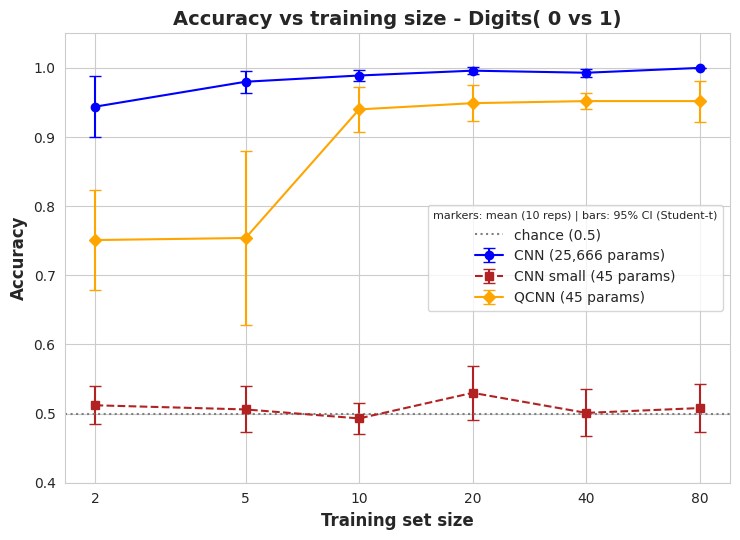}
    \caption{\footnotesize \textbf{Digits ($0$ vs $1$).} QCNN and parameter-matched CNN have $45$ parameters;
    mean over \textbf{10} repetitions, \textbf{50} epochs. The QCNN reaches $\sim$95\% accuracy while the
    matched CNN stays at chance; the $\sim\!570\times$ larger CNN performs best across all sizes.}
    \label{fig:qcnn_vs_cnn_digits}
  \end{subfigure}
  \hfill
  \begin{subfigure}[t]{0.49\textwidth}
    \centering
    \includegraphics[width=\linewidth]{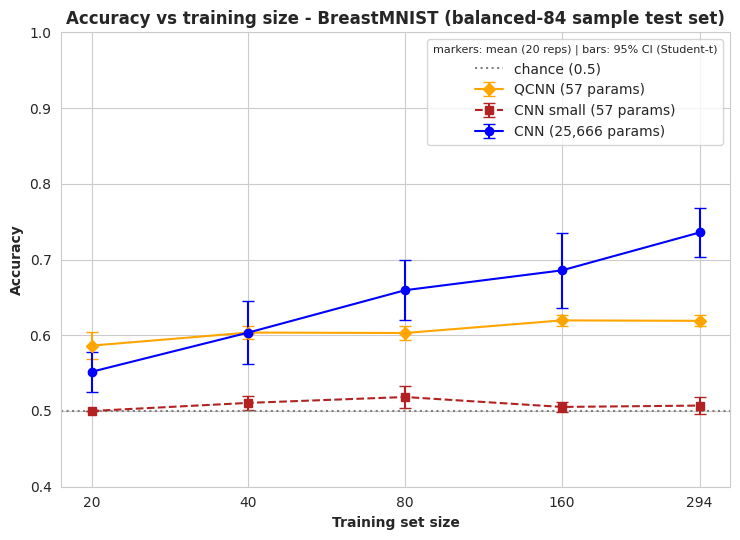}
    \caption{\footnotesize \textbf{BreastMNIST (malignant vs.\ normal/benign).} Balanced $84$-sample test,
    so accuracy $=$ balanced accuracy. QCNN and parameter-matched CNN have $57$ parameters; mean over
    \textbf{20} repetitions, \textbf{100} epochs. The $\sim\!450\times$ larger CNN overtakes the QCNN at larger $N$,
    but in the few-data regime ($N \le 40$) the $57$-parameter QCNN remains competitive (and exceeds it at $N=20$).}
    \label{fig:breastmnist_accuracy_comparison}
  \end{subfigure}
  \caption{\footnotesize Test accuracy versus training set size: QCNN (orange) vs.\ a parameter-matched classical
  CNN (red, same parameter count as the QCNN) and a full classical CNN (blue, $25{,}666$ parameters). Markers are the mean over independent repetitions and error bars denote $95\%$ confidence intervals (Student's $t$ distribution); the dotted line marks chance ($0.5$). In both tasks, at an equal parameter budget the QCNN learns above chance while the parameter-matched CNN
  stays at chance.}
  \label{fig:qcnn_vs_cnn_comparison}
\end{figure*}

\begin{figure*}[t]
  \centering
  \begin{subfigure}[t]{0.49\textwidth}
    \centering
    \includegraphics[width=\linewidth]{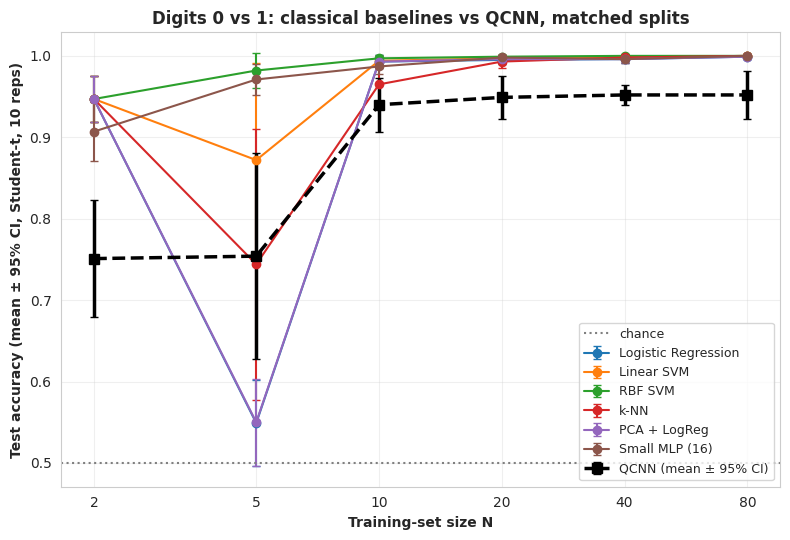}
    \caption{\footnotesize \textbf{Digits ($0$ vs $1$).} QCNN
    (black, dashed) versus a panel of standard classical classifiers; mean over
    $10$ repetitions, error bars are $95\%$ confidence intervals (Student-$t$).
    Simple classical models reach near-perfect accuracy at low capacity and lie
    above the QCNN, which plateaus near $95\%$. Logistic regression and
    PCA${+}$logistic regression nearly coincide.}
    \label{fig:classical_baselines_digits}
  \end{subfigure}
  \hfill
  \begin{subfigure}[t]{0.49\textwidth}
    \centering
    \includegraphics[width=\linewidth]{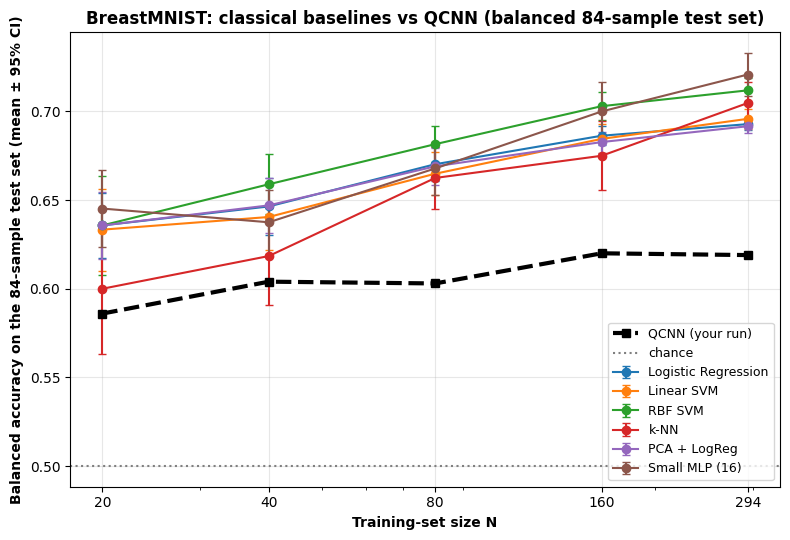}
    \caption{\footnotesize \textbf{BreastMNIST (malignant vs.\
    normal/benign).} Balanced $84$-sample test set, so accuracy $=$ balanced
    accuracy; mean over $20$ repetitions, $95\%$ confidence intervals. Every
    classical baseline lies above the QCNN across all training sizes, the QCNN
    plateauing near $62\%$.}
    \label{fig:classical_baselines_breastmnist}
  \end{subfigure}
  \caption{\footnotesize Test accuracy versus training-set size
  $N$ for the QCNN (black, dashed) and a panel of standard classical baselines,
  logistic regression, linear and RBF support-vector machines, $k$-nearest
  neighbours, PCA${+}$logistic regression, and a small multilayer perceptron,
  evaluated on the identical splits and $\ell_2$-normalized inputs as the QCNN
  (see legend). Markers are the mean over independent repetitions and error bars
  denote $95\%$ confidence intervals (Student-$t$); the dotted line marks chance
  ($0.5$). On both tasks these standard classifiers solve the problem at low
  capacity and lie above the QCNN, confirming that neither task requires a quantum
  model for accuracy. The QCNN's advantage is therefore one of parameter efficiency
  relative to a same-architecture CNN (Fig.~\ref{fig:qcnn_vs_cnn_comparison}), not
  of accuracy relative to classical machine learning.}
  \label{fig:classical_baselines}
\end{figure*}

\subsection{Generalization from few training samples on the digits dataset}

Figure~\ref{fig:digits_result} shows the generalization behaviour of the QCNN on 
the digit task as the training-set size $N$ is varied. For every $N$, training accuracy rises to near unity within a few epochs, while the test metrics depend strongly on $N$: as the training set grows, test loss decreases, test accuracy 
increases, and the gap between training and test loss narrows. The model already learns a useful decision rule in the few-shot regime, with test accuracy well above the $0.5$ chance level at $N=5$ and reaching about $0.94$ by $N=10$, after which it changes little. The confusion matrices in Figure~\ref{fig:cm_digits} show that these accuracy gains are shared across both digit classes rather than driven by a single-class predictor: the number of misclassified test images falls sharply as $N$ grows from the smallest sizes
to $N=10$ and then changes little, mirroring the accuracy curve. Most directly, the estimated generalization error $\mathrm{gen}(\alpha)=R(\alpha)-\hat{R}_N(\alpha)$ falls by more than an order of magnitude over the range of training sizes, from roughly $4\times10^{-2}$ at the smallest $N$ to about $10^{-3}$ at $N=80$. This is consistent with the bound of Caro \emph{et al.}~\cite{caro2022generalization}, under which a QCNN generalizes from a number of samples governed by its parameter count rather than by the dimension of the encoded state space.
We stress, however, that this experiment varies only the training-set size $N$, while the circuit complexity $T$, the degree of parameter reuse $M$, and the architecture are held fixed. A train--test gap that narrows with increasing $N$ is a generic feature of learning models, so these data are qualitatively consistent with, but do not by themselves constitute a test of, the specific dependence on $T$ and $M$ in Eqs.~(4)--(5). A direct test would vary the circuit depth, parameter count, or degree of parameter sharing at fixed $N$ and check the predicted $\sqrt{T}$ and $\log M$ scaling; we identify this as an important direction for future work.

The generalization-error curve is not strictly monotonic: the value at $N=40$ is slightly larger than at $N=20$ and $N=80$. We attribute this to finite-sampling effects, specifically shot noise, random weight initialization, and the limited 
number of repetitions, rather than to a breakdown of the small-data picture. The dominant trend is unambiguous: beyond the extreme few-shot regime, optimization becomes more stable and test performance more reliable as $N$ increases.

These results also broaden the setting in which such behaviour has been observed. The numerical demonstration of Caro \emph{et al.} concerned quantum-phase recognition on quantum states, whereas the present task is binary classification of classical images loaded into the circuit by amplitude encoding. The experiment therefore does not merely reproduce the earlier phase-recognition result; it shows that the same qualitative few-sample behaviour arises in a hardware-compatible QCNN applied to classical image data.

\subsection{BreastMNIST: learning under a strict parameter budget}

The BreastMNIST task is substantially harder and more clinically realistic than the digit benchmark, and absolute performance is correspondingly lower. This lower ceiling is consistent with the structure of the data: the malignant and normal/benign classes overlap substantially in pixel space, in contrast to the cleanly separated digit
classes (Figure~\ref{fig:tsne}). Even so, the QCNN learns consistently above chance (Figure~\ref{fig:breastmnist_result}). On the fixed, class-balanced $84$-sample test set, for which raw accuracy equals balanced 
accuracy, the aggregate balanced accuracy rises from $58.6\%$ at $N=20$ to about 
$62\%$ at the largest training sizes ($62.0\%$ at $N=160$ and $61.9\%$ at $N=294$), passing through $60.4\%$ and $60.3\%$ at $N=40$ and $N=80$. Over the same range, the loss-based generalization gap falls monotonically by roughly a factor of three, from approximately $5.2\times10^{-2}$ at $N=20$ to approximately $1.7\times10^{-2}$ 
at $N=294$, so that larger training sets improve reliability even where the gains in balanced accuracy are modest. Together these trends indicate that the QCNN is not memorizing the training data but extracting a weak yet genuine class signal from a clinically motivated dataset under severe few-shot constraints.

The per-class behaviour supports this reading. Across all training sizes the QCNN retains a balanced error profile rather than collapsing onto a single-class predictor: both classes are recalled above chance, with recall of the malignant class between $54\%$ and $60\%$ and of the normal/benign class between $61\%$ and $65\%$. Because the test set is balanced, a constant classifier would score only $50\%$; the observed balanced accuracy of roughly $59\%$--$62\%$ therefore represents a small but consistent improvement over chance.

\subsection{Comparison with classical baselines}

To isolate the role of model capacity, we compared the QCNN against two classical convolutional networks: one matched to the QCNN's parameter budget, and an unconstrained network with roughly $25{,}000$ parameters. Both tasks are summarized in Figure~\ref{fig:qcnn_vs_cnn_comparison}.

The clearest effect is at matched capacity. On the digit task, the parameter-matched classical CNN ($45$ parameters) does not exceed chance accuracy 
at any training size, whereas the $45$-parameter QCNN reaches about $0.95$ for $N\ge20$ (Figure~\ref{fig:qcnn_vs_cnn_digits}). The same pattern holds on BreastMNIST, where the parameter-matched CNN stays at chance while the $57$-parameter QCNN learns consistently above it (Figure~\ref{fig:breastmnist_accuracy_comparison}). At an identical and very small parameter budget, the quantum model therefore extracts class-discriminating signal that the classical model of equal size does not.

Against the unconstrained classical CNN the QCNN does not claim a blanket advantage, and the outcome is task and regime dependent. On the digits, the larger CNN is strongest throughout, rising from about $0.92$ at $N=2$ to near-perfect accuracy by $N=80$, as expected from its far greater capacity and the maturity of classical training. On BreastMNIST the picture is more nuanced: in the most data-starved regime the QCNN is competitive with or better than the large CNN, exceeding it at $N=20$ ($58.6\%$ versus about $55\%$) and matching it near $N=40$ ($\sim\!60\%$); beyond this point the larger model's capacity becomes decisive, and it pulls ahead to about $74\%$ at $N=294$ while the QCNN plateaus near $62\%$.

These comparisons are best read not as a claim of quantum advantage but as a controlled study of data efficiency. With two to three orders of magnitude fewer parameters than the unconstrained baseline, the QCNN learns where an equally small classical model fails, and it remains competitive with a much larger classical model precisely where data are scarcest; its limited capacity, however, prevents it from exploiting additional data as effectively as the large network, once enough is available.

The parameter-matched comparison above is architecture-matched:
it contrasts the QCNN with a classical convolutional network of the same
convolution-pooling structure and the same parameter budget. To place both in the context of standard classical machine learning, and to confirm that the matched CNN's failure reflects the difficulty of learning at an extremely small CNN capacity rather than a limitation of classical methods in general, we also
evaluated a panel of conventional classifiers on the identical splits and
$\ell_2$-normalized inputs: logistic regression, linear and RBF support-vector
machines, $k$-nearest neighbours, PCA followed by logistic regression, and a small multilayer perceptron (Figure~\ref{fig:classical_baselines}). On the digit task these models solve the problem readily, with an RBF SVM exceeding $0.98$ accuracy by $N=5$ and logistic regression exceeding $0.99$ by $N=10$, both well above the
QCNN. The digit task is therefore easily separable by modest-capacity classical
models, and our comparison should be read accordingly: the QCNN's advantage is one
of parameter efficiency relative to a same-architecture classical network, not of
accuracy relative to classical machine learning.

\section{Scaling behavior\label{sec:scaling}}

\begin{figure*}[t]
  \centering
  \includegraphics[width=\textwidth]{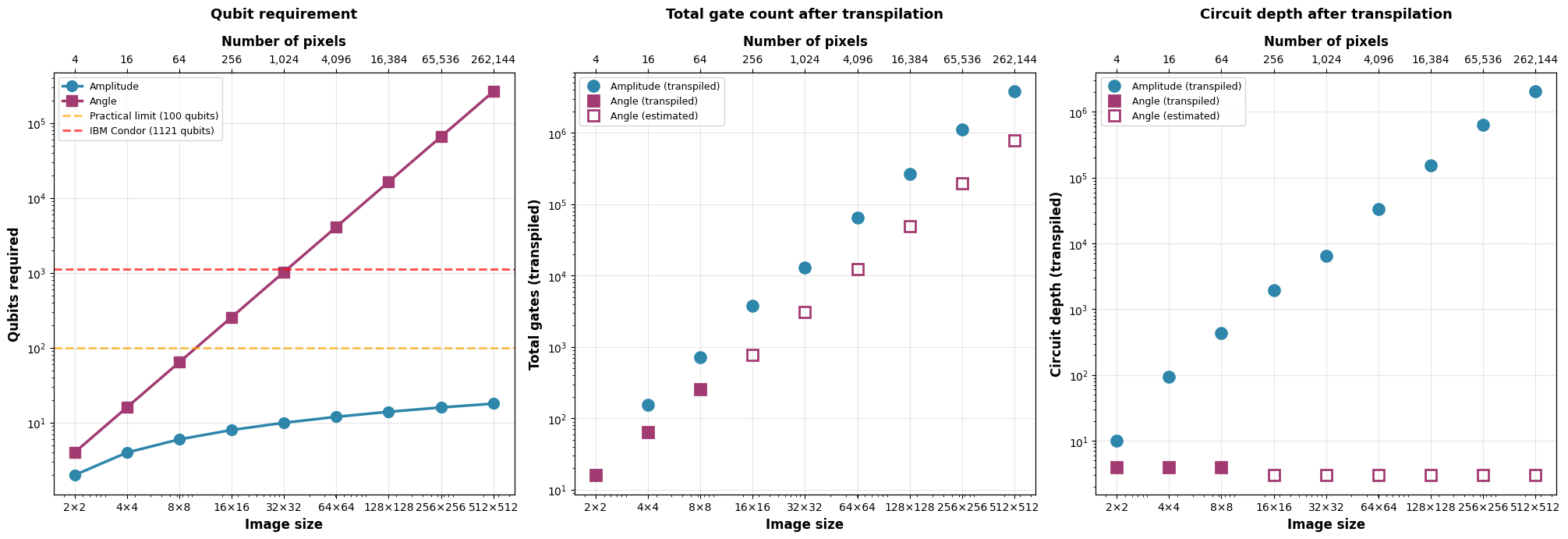}
  \caption{\textbf{Scaling and hardware feasibility of quantum image-encoding circuits.}
  Amplitude (blue) and angle (purple) encoding are compared across image resolutions from
  \(2\times2\) to \(512\times512\) (lower axis: image size; upper axis: corresponding number of pixels).
  All circuits are transpiled to IBM's \texttt{FakeKyiv} backend, a noise-calibrated simulator of the
  \(127\)-qubit IBM~Kyiv quantum processor.
  \textbf{Left:} qubit requirement. Amplitude encoding needs only \(\lceil\log_2 N\rceil\) qubits and stays
  well within current hardware (dashed lines: a \(100\)-qubit practical limit and the \(1121\)-qubit IBM Condor),
  whereas angle encoding needs one qubit per pixel and exceeds these limits beyond small images. \textbf{Center:} total gate count after transpilation. Amplitude encoding incurs a substantially larger gate count due to its multi-controlled state preparation, while angle encoding remains comparatively cheap (filled markers: actual transpilation; open markers: theoretical estimate beyond the directly simulable qubit range). \textbf{Right:} transpiled circuit depth. Amplitude encoding grows deep with resolution, whereas angle encoding
  stays shallow (near-constant depth, as the single-qubit rotations execute in parallel). Overall, the results reveal a fundamental asymmetry: amplitude encoding is \emph{qubit-efficient but depth- and gate-prohibitive}, while angle encoding is \emph{depth-efficient but qubit-prohibitive} for large-scale images.}
  \label{fig:encoding_scaling_top3}
\end{figure*}

\begin{figure*}[t]
  \centering
  \includegraphics[width=\textwidth]{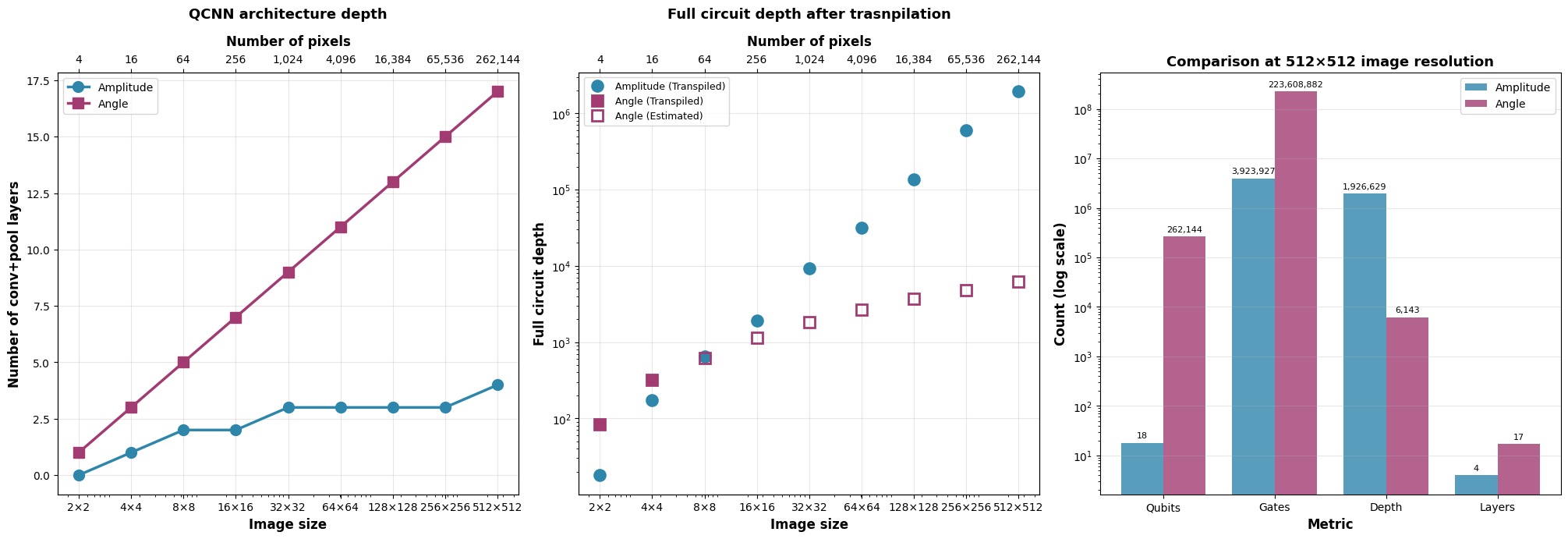}
  \caption{\textbf{Scaling of the full QCNN (encoding $+$ convolution/pooling) under amplitude and angle encoding.}
  Circuits use true mid-circuit-measurement pooling and are transpiled to IBM's \texttt{FakeKyiv} backend across image
  resolutions from \(2\times2\) to \(512\times512\) (lower axis: image size; upper axis: corresponding number of pixels).
  \textbf{Left:} QCNN architecture depth (number of convolution and pooling layers). Because angle encoding starts from one
  qubit per pixel, the layer count grows with the pixel count (\(17\) at \(512\times512\)); amplitude encoding starts
  from only \(\lceil\log_2 P\rceil\) qubits and therefore needs far fewer layers (\(4\)).
  \textbf{Center:} total transpiled circuit depth. Amplitude encoding grows steeply with resolution due to its
  multi-controlled state preparation, whereas angle encoding stays shallow (its single-qubit rotations and pooling
  execute largely in parallel); filled markers denote actual transpilation and open markers a theoretical estimate
  beyond the directly simulable qubit range.
  \textbf{Right:} resource comparison at \(512\times512\). Amplitude encoding requires far fewer qubits
  (\(18\) vs.\ \(262{,}144\)), gates (\(3.9{\times}10^{6}\) vs.\ \(2.2{\times}10^{8}\)), and layers (\(4\) vs.\ \(17\)),
  but a substantially larger circuit depth (\(1.9{\times}10^{6}\) vs.\ \(6{,}143\)).
  Note that the gate-count ordering is reversed relative to the encoding-only analysis
  (Fig.~\ref{fig:encoding_scaling_top3}): there amplitude encoding is the more gate-heavy of the two, whereas in the
  full QCNN angle encoding dominates because its convolution layers act on one qubit per pixel, indicating that this crossover
  originates from the QCNN layers, not from the encoding step alone.
  Overall, amplitude encoding is the only hardware-feasible option at scale with qubit, gate, and layer efficiency, at the
  cost of deep state-preparation circuits, whereas angle encoding remains shallow but qubit-prohibitive.}
  \label{fig:qcnn_scaling}
\end{figure*}

\begin{figure*}[t]
  \centering
  \includegraphics[width=\textwidth]{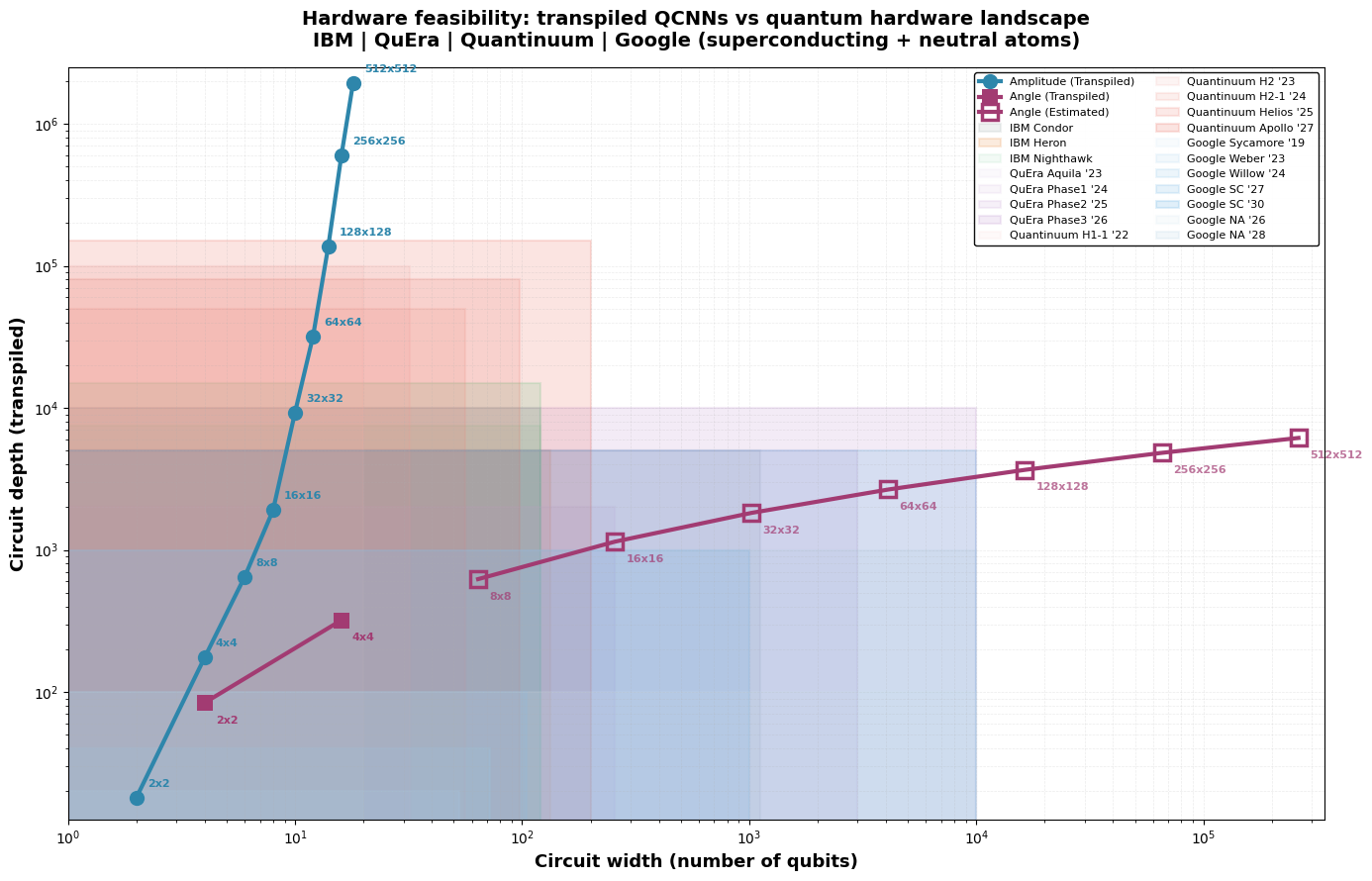}
  \caption{
  \textbf{Hardware feasibility of transpiled QCNN circuits compared to current and projected quantum hardware envelopes across multiple platforms.}
  The figure plots circuit depth (after transpilation) versus circuit width (number of qubits) for QCNN implementations using amplitude and angle encoding across increasing image resolutions (2$\times$2 to 512$\times$512).
  Shaded regions indicate the approximate width--depth capabilities of leading quantum hardware platforms: IBM superconducting processors (Condor, Heron, Nighthawk), QuEra neutral-atom systems (Aquila and the 2024--2026 roadmap), Quantinuum trapped-ion systems (H1, H2, Helios, Apollo), and Google's dual-modality roadmap spanning superconducting (Sycamore, Weber, Willow, and projected 2027--2030 systems) and neutral-atom (2026--2028) architectures.
  For each encoding, filled markers denote circuits small enough to be explicitly transpiled to a hardware-native gate set, while hollow markers denote larger circuits whose resource requirements are estimated beyond the transpilable range.
  Amplitude encoding remains within feasible qubit limits across all tested resolutions but exhibits rapidly increasing circuit depth, approaching or exceeding realistic execution limits for large images.
  In contrast, angle encoding maintains comparatively shallow circuits but requires a rapidly growing number of qubits, exceeding hardware limits at moderate resolutions.
  Overall, the results highlight a fundamental trade-off between qubit efficiency and circuit depth: amplitude encoding is hardware-feasible in width but constrained by depth, whereas angle encoding is depth-efficient but qubit-prohibitive for large-scale quantum image processing.
  The diversity of platform envelopes further illustrates how different hardware modalities (superconducting, trapped-ion, and neutral-atom) offer complementary regions of feasibility for the two encoding strategies.
  }
  \label{fig:hardware_feasibility}
\end{figure*}

The experiments above show that, at the resolutions used for training, optimization and generalization are not the limiting factors. We now ask what constrains the scaling of such a QCNN pipeline to higher-resolution images, and find that the 
binding cost is the data-encoding step rather than the variational layers. We quantify this by transpiling both the encoding circuit alone 
(Figure~\ref{fig:encoding_scaling_top3}) and the full QCNN (Figure~\ref{fig:qcnn_scaling}) to IBM's \texttt{FakeKyiv} backend across image resolutions from $2\times2$ to $512\times512$ pixels, and by placing the resulting circuits against the width-depth envelopes of current and announced hardware 
(Figure~\ref{fig:hardware_feasibility}).

The two encodings occupy opposite extremes of a single trade-off. Amplitude encoding (QPIE) uses only $\lceil\log_2 P\rceil$ qubits for a $P$-pixel image, so even the largest case considered remains narrow: at $512\times512$ the full QCNN 
occupies $18$ qubits, well within current device widths (Figure~\ref{fig:encoding_scaling_top3}, left). This width efficiency is paid for in 
depth. Because preparing an arbitrary amplitude state requires $\mathcal{O}(2^n)$ multi-controlled operations, the transpiled circuit grows steeply with resolution, reaching approximately $3.9\times10^{6}$ gates and a depth of about $1.9\times10^{6}$ at $512\times512$ (Figure~\ref{fig:qcnn_scaling}, right). Amplitude encoding is therefore width-feasible but depth-limited.

Angle encoding shows the opposite pattern. Mapping each pixel to a single-qubit rotation keeps the circuit shallow and almost resolution-independent in depth, but the qubit count grows linearly with the pixel count, reaching $262{,}144$ qubits at 
$512\times512$, which is far beyond any near-term platform even though the corresponding 
transpiled depth ($\sim\!6\times10^{3}$) is orders of magnitude smaller than in the amplitude-encoded case. Angle encoding is thus depth-feasible but width-prohibitive. A concurrent comparative study of QCNN encodings reaches a
compatible conclusion from the performance side: evaluating angle, amplitude, and a hybrid phase/angle encoding under depolarizing noise on downsampled MNIST and Fashion-MNIST, it finds the best encoding to be regime-dependent, with angle encoding favored at the smallest resolutions and amplitude encoding strongest at full
resolution~\cite{feng2025comparative}. Our analysis is complementary: rather than
accuracy at $4\times4$ and $8\times8$, we quantify the qubit, gate, and depth budget
of amplitude and angle encoding across resolutions up to $512\times512$ after
transpilation to hardware.

A further point is that the gate-count ordering reverses between the encoding step and the full pipeline. For the encoding circuit alone, amplitude preparation is the more gate-heavy of the two (Figure~\ref{fig:encoding_scaling_top3}, center); once the convolutional and pooling layers are included, angle encoding instead dominates the 
gate count ($2.2\times10^{8}$ versus $3.9\times10^{6}$ at $512\times512$), because 
its layers act on one qubit per pixel. The crossover therefore originates in the QCNN layers, not in the encoding step alone. Set against hardware envelopes (Figure~\ref{fig:hardware_feasibility}), the two strategies occupy complementary but presently infeasible regions: amplitude-encoded circuits lie within reach in width yet exceed realistic depth budgets at high resolution, whereas angle-encoded 
circuits stay shallow but leave the accessible width range already at moderate resolution. Of the two, amplitude encoding is the only option that remains plausible in width across all tested resolutions, but it is not yet near-term practical 
because of the depth explosion in state preparation.

These results point to encodings of intermediate granularity as the most practical route to scalability. Rather than mapping an entire image either into $\lceil\log_2 P\rceil$ amplitudes or into one qubit per pixel, patch-based and 
hybrid schemes encode small groups of pixels into small registers, trading part of the qubit efficiency of amplitude encoding for a large reduction in state-preparation depth~\cite{balewski2024quantum,lang2024representation}. Such schemes would additionally endow the pipeline with a spatial locality that full-image amplitude encoding does not possess: each patch is encoded independently into a small register and processed hierarchically, so that operations act on localized receptive fields, closer to a classical convolutional network. A systematic determination of where, along this spectrum, a QCNN pipeline first becomes hardware-feasible is left to future work; the present analysis identifies the relevant axis, which is not learning but the qubit-versus-depth cost of loading classical data.

\section{Discussion and conclusion\label{sec:discussion}}

We set out to separate three questions that are often conflated for QCNNs: whether
a hardware-compatible model generalizes from few samples, where the practical
scaling barrier lies, and whether parameter-efficient learning amounts to a quantum
advantage. Our results give a consistent answer. On both a synthetic digit task and
the medically motivated BreastMNIST benchmark, the QCNN generalizes from very small training sets, with a generalization error that decreases with sample size, qualitatively consistent with the bound of Caro et al.~\cite{caro2022generalization}. Because we vary only the
training-set size, this consistency does not by itself test the bound's dependence on circuit complexity and parameter reuse, which we leave as an open experimental question. The dominant obstacle to scaling such a model is not optimization or generalization but the cost of encoding classical data, with amplitude encoding qubit-frugal yet prohibitively
deep and angle encoding shallow yet qubit-prohibitive. And although the QCNN learns where an equally small classical network cannot, it does not surpass an
unconstrained classical CNN, so we read our results as a controlled study of data
efficiency rather than as evidence of quantum advantage.

This framing is most relevant in the settings that motivated the study. In medical
imaging and rare-disease analysis, labelled data are scarce, costly, and
constrained by privacy~\cite{varoquaux2022machine,schafer2024overcoming}, and
methods are increasingly judged by their behaviour in the few-shot regime rather
than by asymptotic accuracy~\cite{pachetti2024systematic}. BreastMNIST is a
deliberately small, balanced proxy for this regime, and the QCNN's ability to
extract weak but reproducible signal from tens of examples, with a parameter budget
two to three orders of magnitude below the classical baseline, is encouraging for
applications where both model size and data are limited.

Our conclusions are consistent with, and complementary to, the concurrent benchmark
of Singh \emph{et al.}~\cite{singh2026benchmarking}, who evaluated pure quantum models on eight MedMNIST datasets, including BreastMNIST, on a $127$-qubit IBM processor with error suppression and mitigation. They report a favourable accuracy-per-parameter profile, classifying with on the order of $10^{2}$ parameters at accuracies ($\sim\!67\%$) and AUCs ($\sim\!72\%$) comparable to classical networks with $10^{4}$ parameters, and describe an advantage over ResNet baselines restricted to the same downsampled inputs, while acknowledging that their models do not outperform the best classical methods. Two differences are instructive. Their study used angle encoding at small qubit counts ($4$-$11$ qubits), which our scaling analysis identifies as precisely the depth-efficient regime that remains executable on present hardware, explaining why a real-hardware demonstration is feasible there, whereas the amplitude-encoded circuits required for higher-resolution inputs are depth-limited. Conversely, that study fixes the training set and does not examine few-shot generalization or the encoding-driven scaling bottleneck, which are the central questions here. The two works thus probe different axes of the same problem: breadth and real-hardware execution on one side, sample complexity and encoding scalability on the other.

Interpreting strong few-shot performance as quantum advantage would be unwarranted
on theoretical grounds. Gil-Fuster \emph{et al.}~\cite{gil2024understanding} showed
that uniform generalization bounds do not by themselves explain QML generalization,
and Bermejo \emph{et al.}~\cite{bermejo2026quantum} showed that QCNNs are
effectively classically simulable on the locally easy benchmarks where they perform
well. More broadly, Cerezo \emph{et al.}~\cite{cerezo2025simulability} argued that
the same structural property that makes QCNNs trainable, the provable absence of
barren plateaus~\cite{pesah2021absence}, often also renders their loss landscapes
classically simulable, and Huang \emph{et al.}~\cite{huang2021power} established
that access to data can make classical models competitive even on tasks of quantum
origin. Our datasets are low-dimensional and highly structured and therefore lie
squarely in the regime these results describe, so strong few-shot performance here
should not be read as a uniquely quantum effect. Where a genuine, even exponential,
advantage may exist is a different regime: Zhao \emph{et al.}~\cite{zhao2026exponential}
show that a polylogarithmically sized fault-tolerant quantum computer can classify
and compress \emph{massive} classical datasets while circumventing the data-loading
bottleneck, exactly the bottleneck our NISQ-era analysis finds binding. Recent
overviews map this landscape of established, conditional, and absent
advantages~\cite{chang2025primer,zimboras2025myths}, and we place our contribution
in its empirical, advantage-agnostic part.

Two caveats temper this reading. First, the classical simulability of QCNNs is
established only for specific regimes: bounded-entanglement inputs, shallow and
structured circuits, and the ``locally easy'' benchmarks on which such models
perform well~\cite{bermejo2026quantum,cerezo2025simulability}. It is not a universal
property. Efficient classical simulation is governed by circuit entanglement, depth, and data complexity rather than by qubit count alone, and as these grow the
simulation is no longer guaranteed to be efficient and may become intractable; the
boundary between the simulable and classically hard regimes, rather than system size per se, is what will determine where a genuinely quantum implementation is warranted. Second, classical simulability precludes a computational quantum advantage but not a practical one: even where a QCNN can be reproduced on a classical computer, its distinctive inductive bias and parameter economy, learning from few samples with few parameters, may confer modelling benefits in data efficiency, model
compactness, and trainability that are independent of any hardware speed-up and
remain to be systematically characterized. Spectral methods~\cite{schuld2021effect, belis2026spectralmethodscrucialmachine} appear to be a route for further investigations.

A fully equivalent QCNN-CNN comparison remains an open problem. Classical CNNs
benefit from decades of architectural and optimization refinement and from
inductive biases tailored to images, whereas QCNNs are early-stage and differ in
capacity, parameter semantics, and trainable structure. We controlled the
comparison as tightly as possible, matching parameter counts, training budgets,
repetitions, and evaluation protocol, and the parameter-matched baseline gives the
cleanest contrast: at an identical budget the QCNN learns while the classical model
does not. Even so, matched parameter count is only one notion of fairness, and
capacity, expressivity, and optimization difficulty are not equalized by it.
Standardized QML benchmarking protocols, with matched architectures and consistent
evaluation pipelines, would make such comparisons more conclusive.

Several limitations qualify these conclusions. The experiments are simulated rather
than executed on hardware: the digit task uses a noise-calibrated dynamic-circuit
model and the BreastMNIST task an idealized statevector backend, so device effects
beyond the \texttt{FakeKyiv} noise model are not captured, and a real-hardware
demonstration of the kind reported by Singh \emph{et al.} remains future work.
Absolute performance on BreastMNIST is modest, and the encoding bottleneck we
identify is quantified rather than resolved. These caveats bound the strength of the
conclusions while leaving the qualitative findings, few-shot generalization and an
encoding-limited scaling, intact.

The most consequential open problem is the encoding bottleneck itself, for which intermediate, patch-based schemes that trade qubit count against state-preparation depth are the natural avenue.  

Beyond image classification, the data efficiency observed here is most valuable where data are intrinsically scarce. Two such directions motivate our ongoing work: integrating a QCNN value estimator into geometric deep learning models for molecular engineering of RNA-based constructs~\cite{bibekar2025context}, and predicting the co-milling dissolution enhancement of poorly water-soluble drugs from molecular descriptors on a dataset of only $29$ compounds~\cite{patzmann2024predictive}. Both lie in the small-data, modest-dimensionality regime in which the inductive bias and parameter economy of
QCNNs are most likely to matter, and both will require the hardware-aware encodings
this study identifies as the binding constraint.

In summary, we provide empirical evidence that a hardware-compatible QCNN
generalizes from very few training samples in a classical image-classification setting, extending the small-data picture of Caro \emph{et al.} from quantum-phase recognition to classical images, and we show that for image-based QML in the NISQ era the dominant barrier is the cost of encoding and executing data at scale rather
than statistical generalization. On a medically relevant benchmark the QCNN does
not surpass an unconstrained classical model, yet it learns meaningful signal with orders of magnitude fewer parameters. We therefore regard QCNNs as useful models for studying data-efficient learning, while stressing that a demonstration of practical advantage will require better encodings, real-hardware execution, stronger application-driven benchmarks, and more rigorous classical baselines.

\section{Data availability}\label{Data}

The results data (training logs and confusion matrices) generated in this study are openly available on Zenodo at
\href{https://doi.org/10.5281/zenodo.22820877}{doi:10.5281/zenodo.22820877}.
The image datasets analyzed are publicly available: the handwritten digits through scikit-learn and BreastMNIST through the MedMNIST collection.

\section{Declaration of AI use}
In preparing this manuscript, AI tools were used only in a supporting role. Their use was confined to editing the language of text that the authors had already written, reorganizing and reformatting content, searching for and checking references, helping to write and debug the analysis and plotting scripts, and cross-checking numerical and symbolic results as well as the internal consistency of the manuscript. The conception of the research, the writing, and the generation and independent verification of every scientific result were done by the authors, who accept full responsibility for the work
in its entirety. This assistance was provided by Claude Code (Anthropic) using
the Claude Opus 4.8 model.

\section{Acknowledgments}\label{Acknowledgment}

We acknowledge team QubiTO for valuable input that helped refine the model during early development. We acknowledge support from armasuisse Science and Technology (S+T), the Swiss Quantum Initiative (SQI) of the Swiss Academy of Sciences (SCNAT) and the State Secretariat for Education, Research and Innovation (SERI), as well as the National Centre of Competence in Research (NCCR) SPIN, funded by the Swiss National Science Foundation (grant number 51NF40-180604).

\newpage

\bibliography{bibliography}

@article{caro2022generalization,
author={Caro, Matthias C. and Huang, Hsin-Yuan and Cerezo, M. and Sharma, Kunal and Sornborger, Andrew and Cincio, Lukasz and Coles, Patrick J.},
title={Generalization in quantum machine learning from few training data},
journal={Nature Communications}, year={2022}, month={Aug}, day={22},
volume={13}, number={1}, pages={4919}, issn={2041-1723},
doi={10.1038/s41467-022-32550-3},
url={https://doi.org/10.1038/s41467-022-32550-3}
}

@online{qml_generalization_tutorial,
  author={Korbinian Kottmann and Luis Mantilla Calderon and Maurice Weber},
  title={Generalization in QML from Few Training Data},
  year={2022},
  url={https://pennylane.ai/qml/demos/tutorial_learning_few_data},
  note={Last updated: 2024-06-08}, organization={PennyLane}
}

@article{gil2024understanding,
  title={Understanding quantum machine learning also requires rethinking generalization},
  author={Gil-Fuster, Elies and Eisert, Jens and Bravo-Prieto, Carlos},
  journal={Nature Communications}, volume={15}, number={1}, pages={2277}, year={2024},
  publisher={Nature Publishing Group UK London},
  doi={10.1038/s41467-024-45882-z},
  url={https://doi.org/10.1038/s41467-024-45882-z}
}

@article{wei2023quantum,
  title={Quantum machine learning in medical image analysis: A survey},
  author={Wei, Lin and Liu, Haowen and Xu, Jing and Shi, Lei and Shan, Zheng and Zhao, Bo and Gao, Yufei},
  journal={Neurocomputing}, volume={525}, pages={42--53}, year={2023}, publisher={Elsevier},
  doi={10.1016/j.neucom.2023.01.049},
  url={https://doi.org/10.1016/j.neucom.2023.01.049}
}

@article{yan2024review,
  title={Review of medical image processing using quantum-enabled algorithms},
  author={Yan, Fei and Huang, Hesheng and Pedrycz, Witold and Hirota, Kaoru},
  journal={Artificial Intelligence Review}, volume={57}, number={11}, pages={300}, year={2024}, publisher={Springer},
  doi={10.1007/s10462-024-10932-x},
  url={https://doi.org/10.1007/s10462-024-10932-x}
}

@article{lang2024representation,
  title={Representation of Classical Data on Quantum Computers},
  author={Lang, Thomas and Heim, Anja and Dremel, Kilian and Prjamkov, Dimitri and Blaimer, Martin and Firsching, Markus and Papadaki, Anastasia and Kasperl, Stefan and Fuchs, Theobald OJ},
  journal={arXiv preprint arXiv:2410.00742}, year={2024},
  doi={10.48550/arXiv.2410.00742},
  url={https://doi.org/10.48550/arXiv.2410.00742}
}

@article{cong2019quantum,
  title={Quantum convolutional neural networks},
  author={Cong, Iris and Choi, Soonwon and Lukin, Mikhail D},
  journal={Nature Physics}, volume={15}, number={12}, pages={1273--1278}, year={2019},
  publisher={Nature Publishing Group UK London},
  doi={10.1038/s41567-019-0648-8},
  url={https://doi.org/10.1038/s41567-019-0648-8}
}

@article{hur2024understanding,
  title={Understanding Generalization in Quantum Machine Learning with Margins},
  author={Hur, Tak and Park, Daniel K},
  journal={arXiv preprint arXiv:2411.06919}, year={2024},
  doi={10.48550/arXiv.2411.06919},
  url={https://doi.org/10.48550/arXiv.2411.06919}
}

@article{banchi2021generalization,
  title={Generalization in quantum machine learning: A quantum information standpoint},
  author={Banchi, Leonardo and Pereira, Jason and Pirandola, Stefano},
  journal={PRX Quantum}, volume={2}, number={4}, pages={040321}, year={2021}, publisher={APS},
  doi={10.1103/PRXQuantum.2.040321},
  url={https://doi.org/10.1103/PRXQuantum.2.040321}
}

@article{balewski2024quantum,
  title={Quantum-parallel vectorized data encodings and computations on trapped-ion and transmon QPUs},
  author={Balewski, Jan and Amankwah, Mercy G and Van Beeumen, Roel and Bethel, E Wes and Perciano, Talita and Camps, Daan},
  journal={Scientific Reports}, volume={14}, number={1}, pages={3435}, year={2024},
  publisher={Nature Publishing Group UK London},
  doi={10.1038/s41598-024-53720-x},
  url={https://doi.org/10.1038/s41598-024-53720-x}
}

@article{carrera2024combining,
  title={Combining quantum processors with real-time classical communication},
  author={Carrera Vazquez, Almudena and Tornow, Caroline and Rist{\`e}, Diego and Woerner, Stefan and Takita, Maika and Egger, Daniel J},
  journal={Nature}, volume={636}, pages={75--79}, year={2024},
  publisher={Nature Publishing Group UK London},
  doi={10.1038/s41586-024-08178-2},
  url={https://doi.org/10.1038/s41586-024-08178-2}
}

@article{tripp2024measuring,
  title={Measuring the energy consumption and efficiency of deep neural networks: An empirical analysis and design recommendations},
  author={Tripp, Charles Edison and Perr-Sauer, Jordan and Gafur, Jamil and Nag, Amabarish and Purkayastha, Avi and Zisman, Sagi and Bensen, Erik A},
  journal={arXiv preprint arXiv:2403.08151}, year={2024},
  doi={10.48550/arXiv.2403.08151},
  url={https://doi.org/10.48550/arXiv.2403.08151}
}

@article{zimboras2025myths,
  title={Myths around quantum computation before full fault tolerance: What no-go theorems rule out and what they don't},
  author={Zimbor{\'a}s, Zolt{\'a}n and Koczor, B{\'a}lint and Holmes, Zo{\"e} and Borrelli, Elsi-Mari and Gily{\'e}n, Andr{\'a}s and Huang, Hsin-Yuan and Cai, Zhenyu and Ac{\'\i}n, Antonio and Aolita, Leandro and Banchi, Leonardo and others},
  journal={arXiv preprint arXiv:2501.05694}, year={2025},
  doi={10.48550/arXiv.2501.05694},
  url={https://doi.org/10.48550/arXiv.2501.05694}
}

@article{zhao2026exponential,
  title={Exponential quantum advantage in processing massive classical data},
  author={Zhao, Haimeng and Zlokapa, Alexander and Neven, Hartmut and Babbush, Ryan and Preskill, John and McClean, Jarrod R and Huang, Hsin-Yuan},
  journal={arXiv preprint arXiv:2604.07639}, year={2026},
  doi={10.48550/arXiv.2604.07639},
  url={https://doi.org/10.48550/arXiv.2604.07639}
}

@article{bermejo2026quantum,
  title={Quantum convolutional neural networks are effectively classically simulable},
  author={Bermejo, Pablo and Braccia, Paolo and Rudolph, Manuel S and Holmes, Zo{\"e} and Cincio, Lukasz and Cerezo, Marco},
  journal={PRX Quantum}, volume={7}, number={2}, pages={020304}, year={2026}, publisher={APS},
  doi={10.1103/8qt9-72ts},
  url={https://doi.org/10.1103/8qt9-72ts}
}

@article{biamonte2017quantum,
  title={Quantum machine learning},
  author={Biamonte, Jacob and Wittek, Peter and Pancotti, Nicola and Rebentrost, Patrick and Wiebe, Nathan and Lloyd, Seth},
  journal={Nature}, volume={549}, number={7671}, pages={195--202}, year={2017},
  publisher={Nature Publishing Group UK London},
  doi={10.1038/nature23474},
  url={https://doi.org/10.1038/nature23474}
}

@article{cerezo2021variational,
  title={Variational quantum algorithms},
  author={Cerezo, Marco and Arrasmith, Andrew and Babbush, Ryan and Benjamin, Simon C and Endo, Suguru and Fujii, Keisuke and McClean, Jarrod R and Mitarai, Kosuke and Yuan, Xiao and Cincio, Lukasz and Coles, Patrick J},
  journal={Nature Reviews Physics}, volume={3}, number={9}, pages={625--644}, year={2021},
  publisher={Nature Publishing Group UK London},
  doi={10.1038/s42254-021-00348-9},
  url={https://doi.org/10.1038/s42254-021-00348-9}
}

@article{huang2021power,
  title={Power of data in quantum machine learning},
  author={Huang, Hsin-Yuan and Broughton, Michael and Mohseni, Masoud and Babbush, Ryan and Boixo, Sergio and Neven, Hartmut and McClean, Jarrod R},
  journal={Nature Communications}, volume={12}, number={1}, pages={2631}, year={2021},
  publisher={Nature Publishing Group UK London},
  doi={10.1038/s41467-021-22539-9},
  url={https://doi.org/10.1038/s41467-021-22539-9}
}

@article{patzmann2024predictive,
  title={Predictive computational models for assessing the impact of co-milling on drug dissolution},
  author={P{\"a}tzmann, N. and O'Dwyer, P. J. and Ber{\'a}nek, J. and Kuentz, M. and Griffin, B. T.},
  journal={European Journal of Pharmaceutical Sciences}, volume={198}, pages={106780}, year={2024}, publisher={Elsevier},
  doi={10.1016/j.ejps.2024.106780},
  url={https://doi.org/10.1016/j.ejps.2024.106780}
}

@article{pesah2021absence,
  title={Absence of barren plateaus in quantum convolutional neural networks},
  author={Pesah, Arthur and Cerezo, M and Wang, Samson and Volkoff, Tyler and Sornborger, Andrew T and Coles, Patrick J},
  journal={Physical Review X}, volume={11}, number={4}, pages={041011}, year={2021},
  publisher={American Physical Society},
  doi={10.1103/PhysRevX.11.041011},
  url={https://doi.org/10.1103/PhysRevX.11.041011}
}

@article{shende2006synthesis,
  title={Synthesis of quantum-logic circuits},
  author={Shende, Vivek V and Bullock, Stephen S and Markov, Igor L},
  journal={IEEE Transactions on Computer-Aided Design of Integrated Circuits and Systems},
  volume={25}, number={6}, pages={1000--1010}, year={2006}, publisher={IEEE},
  doi={10.1109/TCAD.2005.855930},
  url={https://doi.org/10.1109/TCAD.2005.855930}
}

@book{schuld2021machine,
  title={Machine Learning with Quantum Computers},
  author={Schuld, Maria and Petruccione, Francesco},
  series={Quantum Science and Technology}, edition={2}, year={2021},
  publisher={Springer}, address={Cham},
  doi={10.1007/978-3-030-83098-4},
  url={https://doi.org/10.1007/978-3-030-83098-4}
}

@article{varoquaux2022machine,
  title={Machine learning for medical imaging: methodological failures and recommendations for the future},
  author={Varoquaux, Ga{\"e}l and Cheplygina, Veronika},
  journal={npj Digital Medicine}, volume={5}, number={1}, pages={48}, year={2022},
  publisher={Nature Publishing Group UK London},
  doi={10.1038/s41746-022-00592-y},
  url={https://doi.org/10.1038/s41746-022-00592-y}
}

@article{schafer2024overcoming,
  title={Overcoming data scarcity in biomedical imaging with a foundational multi-task model},
  author={Sch{\"a}fer, Raphael and Nicke, Till and H{\"o}fener, Henning and Lange, Annkristin and Merhof, Dorit and Feuerhake, Friedrich and Schulz, Volkmar and Lotz, Johannes and Kiessling, Fabian},
  journal={Nature Computational Science}, volume={4}, number={7}, pages={495--509}, year={2024},
  publisher={Nature Publishing Group US New York},
  doi={10.1038/s43588-024-00662-z},
  url={https://doi.org/10.1038/s43588-024-00662-z}
}

@article{pachetti2024systematic,
  title={A systematic review of few-shot learning in medical imaging},
  author={Pachetti, Eva and Colantonio, Sara},
  journal={Artificial Intelligence in Medicine}, volume={156}, pages={102949}, year={2024}, publisher={Elsevier},
  doi={10.1016/j.artmed.2024.102949},
  url={https://doi.org/10.1016/j.artmed.2024.102949}
}

@article{yao2017quantum,
  title={Quantum image processing and its application to edge detection: theory and experiment},
  author={Yao, Xi-Wei and Wang, Hengyan and Liao, Zeyang and Chen, Ming-Cheng and Pan, Jian and Li, Jun and Zhang, Kechao and Lin, Xingcheng and Wang, Zhehui and Luo, Zhihuang and Zheng, Wenqiang and Li, Jianzhong and Zhao, Meisheng and Peng, Xinhua and Suter, Dieter},
  journal={Physical Review X}, volume={7}, number={3}, pages={031041}, year={2017},
  publisher={American Physical Society},
  doi={10.1103/PhysRevX.7.031041},
  url={https://doi.org/10.1103/PhysRevX.7.031041}
}

@article{scikit-learn,
  title={Scikit-learn: Machine Learning in {P}ython},
  author={Pedregosa, F. and Varoquaux, G. and Gramfort, A. and Michel, V. and Thirion, B. and Grisel, O. and Blondel, M. and Prettenhofer, P. and Weiss, R. and Dubourg, V. and Vanderplas, J. and Passos, A. and Cournapeau, D. and Brucher, M. and Perrot, M. and Duchesnay, E.},
  journal={Journal of Machine Learning Research}, volume={12}, pages={2825--2830}, year={2011},
  url={https://www.jmlr.org/papers/v12/pedregosa11a.html}
}

@article{medmnist,
  title={MedMNIST v2 - A large-scale lightweight benchmark for 2D and 3D biomedical image classification},
  author={Yang, Jiancheng and Shi, Rui and Wei, Donglai and Liu, Zequan and Zhao, Lin and Ke, Bilian and Pfister, Hanspeter and Ni, Bingbing},
  journal={Scientific Data}, volume={10}, number={1}, pages={41}, year={2023},
  publisher={Nature Publishing Group UK London},
  doi={10.1038/s41597-022-01721-8},
  url={https://doi.org/10.1038/s41597-022-01721-8}
}

@article{mitarai2018quantum,
  title={Quantum circuit learning},
  author={Mitarai, Kosuke and Negoro, Makoto and Kitagawa, Masahiro and Fujii, Keisuke},
  journal={Physical Review A}, volume={98}, number={3}, pages={032309}, year={2018},
  publisher={American Physical Society},
  doi={10.1103/PhysRevA.98.032309},
  url={https://doi.org/10.1103/PhysRevA.98.032309}
}

@article{kingma2014adam,
  title={Adam: A Method for Stochastic Optimization},
  author={Kingma, Diederik P. and Ba, Jimmy},
  journal={arXiv preprint arXiv:1412.6980}, year={2014},
  doi={10.48550/arXiv.1412.6980},
  url={https://doi.org/10.48550/arXiv.1412.6980}
}

@article{singh2026benchmarking,
  title={Benchmarking MedMNIST dataset on real quantum hardware},
  author={Singh, Gurinder and Jin, Hongni and Merz Jr., Kenneth M.},
  journal={Scientific Reports}, volume={16}, pages={9017}, year={2026},
  publisher={Nature Publishing Group UK London},
  doi={10.1038/s41598-026-35605-3},
  url={https://doi.org/10.1038/s41598-026-35605-3}
}

@article{cerezo2025simulability,
  title={Does provable absence of barren plateaus imply classical simulability?},
  author={Cerezo, M. and Larocca, Martin and Garc{\'\i}a-Mart{\'\i}n, Diego and Diaz, N. L. and Braccia, Paolo and Fontana, Enrico and Rudolph, Manuel S. and Bermejo, Pablo and Ijaz, Aroosa and Thanasilp, Supanut and Anschuetz, Eric R. and Holmes, Zo{\"e}},
  journal={Nature Communications}, volume={16}, pages={7907}, year={2025},
  publisher={Nature Publishing Group UK London},
  doi={10.1038/s41467-025-63099-6},
  url={https://doi.org/10.1038/s41467-025-63099-6}
}

@misc{chang2025primer,
  title={A Primer on Quantum Machine Learning},
  author={Chang, Su Yeon and Cerezo, M.},
  year={2025}, eprint={2511.15969}, archivePrefix={arXiv}, primaryClass={quant-ph},
  doi={10.48550/arXiv.2511.15969},
  url={https://doi.org/10.48550/arXiv.2511.15969}
}

@article{abriata2026casp16,
  title={Practical Outcomes From {CASP16} for Users in Need of Biomolecular Structure Prediction},
  author={Abriata, Luciano A. and Dal Peraro, Matteo},
  journal={Proteins: Structure, Function, and Bioinformatics},
  volume={94},
  number={1},
  pages={435--446},
  year={2026},
  publisher={Wiley},
  doi={10.1002/prot.70078},
  url={https://doi.org/10.1002/prot.70078}
}

@article{bibekar2025context,
  title={Context-aware geometric deep learning for {RNA} sequence design},
  author={Bibekar, Parth and Krapp, Lucien F. and Dal Peraro, Matteo},
  journal={bioRxiv},
  year={2025},
  publisher={openRxiv},
  doi={10.1101/2025.06.21.660801},
  url={https://doi.org/10.1101/2025.06.21.660801},
  note={Preprint}
}

@article{feng2025comparative,
  title         = {A Comparative Study of Encoding Strategies for Quantum Convolutional Neural Networks},
  author        = {Feng, Xingyun},
  year          = {2025},
  eprint        = {2512.12512},
  archivePrefix = {arXiv},
  primaryClass  = {quant-ph},
  doi           = {10.48550/arXiv.2512.12512}
}

@article{schuld2021effect,
  title   = {Effect of data encoding on the expressive power of variational quantum-machine-learning models},
  author  = {Schuld, Maria and Sweke, Ryan and Meyer, Johannes Jakob},
  journal = {Phys. Rev. A},
  volume  = {103}, number = {3}, pages = {032430}, year = {2021},
  doi     = {10.1103/PhysRevA.103.032430}
}

@article{schreiber2023classical,
  title   = {Classical Surrogates for Quantum Learning Models},
  author  = {Schreiber, Franz J. and Eisert, Jens and Meyer, Johannes Jakob},
  journal = {Phys. Rev. Lett.},
  volume  = {131}, number = {10}, pages = {100803}, year = {2023},
  doi     = {10.1103/PhysRevLett.131.100803}
}

@article{landman2022classically,
  title         = {Classically Approximating Variational Quantum Machine Learning with Random Fourier Features},
  author        = {Landman, Jonas and Thabet, Slimane and Dalyac, Constantin and Mhiri, Hela and Kashefi, Elham},
  year          = {2022},
  eprint        = {2210.13200}, archivePrefix = {arXiv}, primaryClass = {quant-ph},
  doi           = {10.48550/arXiv.2210.13200}
}

@misc{belis2026spectralmethodscrucialmachine,
      title={Spectral methods: crucial for machine learning, natural for quantum computers?}, 
      author={Vasilis Belis and Joseph Bowles and Rishabh Gupta and Evan Peters and Maria Schuld},
      year={2026},
      eprint={2603.24654},
      archivePrefix={arXiv},
      primaryClass={quant-ph},
      url={https://arxiv.org/abs/2603.24654}, 
}

\end{document}